\documentclass[a4paper,12pt]{article}

\usepackage{geometry}
\usepackage{setspace}
\usepackage[hidelinks,linktocpage=true,linktoc=all]{hyperref}
\usepackage{cite}

\usepackage{amsmath}
\usepackage{amssymb}

\newcommand{\affil}{\emph{D\'epartement de Physique Th\'eorique, Universit\'e de Gen\`eve, \\ 24 quai Ernest Ansermet, 1211 Gen\`eve 4, Switzerland}}

\newcommand{\beq}{\begin{equation}}
\newcommand{\eeq}{\end{equation}}
\newcommand{\beqa}{\begin{eqnarray}}
\newcommand{\eeqa}{\end{eqnarray}}

\title{Cosmology in Minkowski space}
\author{Lucas Lombriser}
\date{\small\affil\vspace{2mm}\\\today}

\begin{document}

\onehalfspacing

\maketitle
\thispagestyle{empty}

\abstract{
    Theoretical and observational challenges to standard cosmology such as the cosmological constant problem and tensions between cosmological model parameters inferred from different observations motivate the development and search of new physics.
    A less radical approach to venturing beyond the standard model is the simple mathematical reformulation of our theoretical frameworks underlying it.
    While leaving physical measurements unaffected, this can offer a reinterpretation and even solutions of these problems.
    In this spirit, metric transformations are performed here that cast our Universe into different geometries.
    Of particular interest thereby is the formulation of cosmology in Minkowski space.
    Rather than an expansion of space, spatial curvature, and small-scale inhomogeneities and anisotropies, this frame exhibits a variation of mass, length and time scales across spacetime.
    Alternatively, this may be interpreted as an evolution of fundamental constants.
    As applications of this reframed cosmological picture, the naturalness of the cosmological constant is reinspected and promising candidates of geometric origin are explored for dark matter, dark energy, inflation and baryogenesis.
    An immediate observation thereby is the apparent absence of the cosmological constant problem in the Minkowski frame.
    The formalism is also applied to identify new observable signatures of conformal inhomogeneities, which have been proposed as simultaneous solution of the observational tensions in the Hubble constant, the amplitude of matter fluctuations, and the gravitational lensing amplitude of cosmic microwave background anisotropies.
    These are found to enhance redshifts to distant galaxy clusters and introduce a mass bias with cluster masses inferred from gravitational lensing exceeding those inferred kinematically or dynamically.
}

\newpage
\thispagestyle{empty}

\tableofcontents

\newpage

\section{Introduction} \label{sec:intro}

Its capacity to reproduce the wealth of cosmological observations conducted over the past few decades has cemented $\Lambda$CDM as the standard model of cosmology.
Despite this resounding observational success, significant gaps persist in our theoretical understanding of some of its key ingredients.
The cosmological constant problem~\cite{Weinberg:1988cp, Martin:2012bt}, the nature of dark matter~\cite{Bertone:2016nfn}
and inflation~\cite{Baumann:2009ds, Linde:2014nna}, or the origin of the matter-antimatter asymmetry~\cite{Riotto:1998bt, Canetti:2012zc}
in the Universe constitute particularly important and profound enigmas among them.
The cosmological constant is generally thought to be attributed to the gravitational contribution of vacuum fluctuations, anticipated of adequate magnitude to account for the observed late-time accelerated expansion of the Universe.
Quantum theoretical estimates so far, however, exceed measurement by several orders of magnitude.
Moreover, the comparable size of its associated energy density with that of matter today provokes a further conundrum, the \emph{Why Now?}~or coincidence problem.
The energy density of matter in itself poses a deep physical puzzle.
It is dominated by an elusive dark matter that despite its overwhelming indirect observational evidence has so far evaded direct detection.
The remaining baryonic part, although known in nature, comes with a larger abundance of baryons than antibaryons.
The associated baryogenesis process causing this asymmetry has not been identified yet.
Neither has the precise physical mechanism for an inflationary epoch in the early Universe been determined.

Besides these theoretical shortcomings, a number of unexplained larger and smaller observational tensions have manifested between the cosmological model parameters inferred from our different data sets.
The most prominent among them is the $\sim5\sigma$ tension in the current cosmic expansion rate $H_0$ as measured with the cosmic microwave background (CMB) and with the local distance ladder~\cite{Riess:2021jrx}.
Another $\sim 3\sigma$ discrepancy is observed in the current amplitude of large-scale matter density fluctuations $\sigma_8$ (or $S_8$) as inferred from the CMB and as measured with large-scale structure surveys~\cite{Heymans:2020gsg}.
For another example, a discrepancy of $\lesssim3\sigma$ also manifests in the imprint of weak gravitational lensing on the CMB spectra, which is larger than what would be expected from the cosmological parameters obtained from early-time CMB physics~\cite{Aghanim:2018eyx}.
Interestingly, the preferred enhancement $A_L^{TT}>1$ of the lensing amplitude is not observed in the power spectrum of the reconstructed lensing potential, $A_L^{\phi\phi}=1$, which may be perceived as another indicator of an underlying inconsistency.

Both these theoretical and observational challenges have prompted for the development and search of new physics~\cite{Bull:2015stt, Perivolaropoulos:2021jda}.
In contrast, a less radical approach to venturing beyond the Standard Model of Particle Physics and $\Lambda$CDM is the simple mathematical reformulation of our theoretical frameworks underlying them.
This can offer new perspectives with different physical interpretations and possibly even provide solutions for these theoretical and observational problems.

This article explores the implications of casting cosmology into different spacetime geometries as a simple mathematical manipulation that leaves physical measurements unaffected but can reveal new physical insights.
Clearly, the Minkowski spacetime occupies a special place among the metrics one can transform the cosmic geometry into, being static and flat as well as the spacetime of special relativity and quantum field theory or indeed of the entire Standard Model.
Hence, a particular focus is put on casting cosmology into Minkowski space.
Besides rederiving the Friedmann equations, the evolution of energy densities, and redshift,  the formalism is also employed to gain new perspectives on the cosmological constant problem, dark matter, dark energy, inflation, baryogenesis, and a range of observational cosmic tensions.

The article is organised as follows.
Sec.~\ref{sec:gravity} discusses the general interchangeability of gravitational theory, the energy-momentum sector and spacetime geometry in the Einstein field equations.
Conformal metric transformations and the dimensional reduction of warped spacetimes are briefly reviewed and applied to the Einstein-Hilbert action and the Einstein field equations.
In Sec.~\ref{sec:minkowskicosmology} these transformations are employed to cast cosmology into static and Minkowski space for smooth backgrounds with and without spatial curvature as well as for scalar and tensor perturbations.
The Friedmann equations, the evolution of energy densities, and redshift are derived for the transformed spacetimes.
Conformal transformations are then applied in Sec.~\ref{sec:backgrounds} to cast the cosmic spacetime into Einstein-de Sitter and de Sitter backgrounds as well as for the treatment of nonlinear conformal inhomogeneities.
Sec.~\ref{sec:applications} is devoted to applications of the formalism.
It discusses the naturalness of the cosmological constant in the Minkowski frame, conformal inhomogeneities as solution to tensions between cosmological observations, and promising candidates for dark matter, inflation and baryogenesis.
Finally, conclusions are presented in Sec.~\ref{sec:conclusions}.

\section{Gravity in different casks} \label{sec:gravity}

Before delving into the geometric transformations of cosmological spacetime, the resulting change of physical equations and its implications for their
physical
interpretation and our understanding of the Universe
and the physical processes taking place within it,
a brief presentation of the relevant theoretical background shall be given here.
Sec.~\ref{sec:degeneracy} introduces, as a general concept, the interchangeability of gravitational theory, the energy-momentum sector and spacetime geometry in the Einstein field equations.
Sec.~\ref{sec:transformations} discusses the metric transformations and decomposition that will be used to recast the Einstein-Hilbert action and Einstein field equations into different geometries.
Readers familiar with Riemannian geometry may want to skip this section.
Sec.~\ref{sec:fieldeqs} derives the Einstein-Hilbert action and the field equations under a conformal transformation of the metric, which will be used in Sec.~\ref{sec:minkowskicosmology}.
These discussions are intended to be compact but sufficient to equip the unfamiliar reader with the necessary tools for immersing into the subsequent sections.

\subsection{Interchanging energy-momentum, gravity and geometry} \label{sec:degeneracy}

Let us begin with some general considerations on the relationship between gravitational theory, energy-momentum and geometry in the Einstein field equations, $G_{\mu\nu} = \kappa^2 T_{\mu\nu}$, where $\kappa^2 \equiv 8\pi G c^{-4}$ denotes the gravitational coupling and the metric signature $(-,+,+,+)$ will be adopted throughout the article.
Suppose the gravitational theory is modified, and the energy-momentum tensor $T_{\mu\nu}$ relates to a modified tensor $M_{\mu\nu}$ rather than to the Einstein tensor $G_{\mu\nu}$ such that
\beq
    M_{\mu\nu} = \kappa^2 T_{\mu\nu} \,. \label{eq:mg}
\eeq
This relation can always be recast as the usual Einstein field equation with the definition of an effective energy-momentum tensor,
\beq
    G_{\mu\nu} = \kappa^2 T_{\mu\nu} + \kappa^2 T^{eff}_{\mu\nu} \,, \quad \quad T^{eff}_{\mu\nu} \equiv \kappa^{-2}G_{\mu\nu} - \kappa^{-2}M_{\mu\nu} \,,
\eeq
where covariant conservation of $T_{\mu\nu}$ and the Bianchi identities imply covariant conservation of $T^{eff}_{\mu\nu}$.
This ambiguity between modifications in the metric and matter sectors has also been coined the dark degeneracy in Refs.~\cite{Kunz:2007rk, Lombriser:2015sxa}.

We shall now add another layer of degeneracy, stemming from the choice of geometry.
Let us assume the metric $\hat{g}_{\mu\nu}$ satisfies
\beq
    \hat{G}_{\mu\nu} = \kappa^2 \hat{T}_{\mu\nu} \label{eq:geomdeg}
\eeq
for some specified matter content in $\hat{T}_{\mu\nu}$.
The field equations can now be recast into a different geometry $\tilde{g}_{\mu\nu} = A_{\;\;\;\mu\nu}^{\alpha\beta} \hat{g}_{\alpha\beta} + B_{\mu\nu}$ as
\beq
    \tilde{G}_{\mu\nu} = \kappa^2 \tilde{T}^{eff}_{\mu\nu} \,, \quad \quad \tilde{T}^{eff}_{\mu\nu}(\tilde{g}) \equiv \left(\kappa^{-2}\tilde{G}_{\mu\nu} - \kappa^{-2}\hat{G}_{\mu\nu} + \hat{T}_{\mu\nu}\right)(\tilde{g}) \,. \label{eq:geomdegrecast}
\eeq
This is merely a mathematical manipulation, simply a substitution of the variable one is solving the differential equations in Eq.~\eqref{eq:geomdeg} for.
Just as with any other differential equation, one may always perform such a change of variable, in this case the metric.
The physics remains unchanged.
Note, however, that freely falling particles in $\hat{T}_{\mu\nu}(\tilde{g})$ no longer follow geodesics for $\tilde{g}_{\alpha\beta}$, which manifests as an additional interaction and breaks a degeneracy for visible matter species between the two frames set by the two different metrics.
In contrast, an interacting dark sector is simply receiving additional interactions.
Alternatively to Eq.~\eqref{eq:geomdegrecast}, one could also define modified geometric and matter sectors such as $\tilde{M}_{\mu\nu} \equiv \hat{G}_{\mu\nu}(\tilde{g})$ and $\tilde{T}^{eff}_{\mu\nu} \equiv \hat{T}_{\mu\nu}(\tilde{g})$ in $\tilde{M}_{\mu\nu}(\tilde{g}) = \kappa^2 \tilde{T}^{eff}_{\mu\nu}(\tilde{g})$ or other terms that are Eq.~\eqref{eq:geomdeg} in disguise.
A consequence of these reframed field equations is that one may cast the physics of a system into a spacetime geometry of choice.
The implications of that for our physical interpretation and understanding of the observed Universe will be the main subject of this article.

Note that the transformation between the Jordan and Einstein frames of modified gravity models is related to Eq.~\eqref{eq:geomdegrecast}.
Hereby, a metric transformation is applied to recast the field equations $M_{\mu\nu}(g) = T_{\mu\nu}(g)$ in Jordan frame into Einstein frame with $\tilde{G}_{\mu\nu}(\tilde{g}) = \tilde{T}^{eff}_{\mu\nu}(\tilde{g})$, where $\tilde{T}^{eff}_{\mu\nu}(\tilde{g})$ contains $T_{\mu\nu}(\tilde{g})$ and contributions from $M_{\mu\nu}(\tilde{g})$.
Thereby, $M_{\mu\nu}(g)$ and $T_{\mu\nu}(g)$ as well as $\tilde{G}_{\mu\nu}(\tilde{g})$ and $\tilde{T}^{eff}_{\mu\nu}(\tilde{g})$ are covariantly conserved whereas $M_{\mu\nu}(\tilde{g})$ and $T_{\mu\nu}(\tilde{g})$ are not.

Finally, one may also impose a specific matter content and invert Eq.~\eqref{eq:mg}.
For that, assume the following field equations hold for $g_{\mu\nu}$,
\beq
    G_{\mu\nu} = \kappa^2 T^{(1)}_{\mu\nu} \,.
\eeq
One can now enforce a different matter content $T^{(2)}_{\mu\nu}(g)$ such that
\beq
    M_{\mu\nu} = \kappa^2 T^{(2)}_{\mu\nu} \,, \quad \quad M_{\mu\nu}(g) \equiv \left(G_{\mu\nu} + \kappa^2 T^{(2)}_{\mu\nu} - \kappa^2 T^{(1)}_{\mu\nu}\right)(g) \,. \label{eq:nodark}
\eeq
For example, one may want to evade invoking dark matter and/or dark energy to explain our observations and rather interpret them as some effective components attributed to an underlying modification of gravity.
Eq.~\eqref{eq:nodark} implies that one can in principle construct such a modified gravity model which will be phenomenologically equivalent to $\Lambda$CDM.
It can be obtained from including dark matter and/or dark energy in $T^{(1)}_{\mu\nu}$ but not in $T^{(2)}_{\mu\nu}$.
The intricate part is to find $T^{(2)}_{\mu\nu}$ as a function of $g_{\mu\nu}$ so that the procedure is universal, and when adopting a different metric than $g_{\mu\nu}$ to describe some other gravitating system, the predicted phenomenology of this new system remains invariant, or at least it should remain compatible with the measurements of that.
Of course, for the model to be falsifiable, one would hope that this invariance is broken in some systems.

Instead of specifying the matter content in Eq.~\eqref{eq:nodark}, one could enforce a specific modified tensor $M_{\mu\nu}$ and adapt the energy-momentum tensor accordingly.
One may also specify the matter content and the metric and construct the tensor $M_{\mu\nu}$ valid for that.
Generally, one can specify any two components between (modified) Einstein tensor, energy-momentum tensor and metric by adaption of the third.

The form of the tensors on either side of the field equation~\eqref{eq:mg} is thus not a useful discriminator for the classification between dark sector, interacting dark sector and modified gravity models.
Instead, the distinction can be drawn upon whether the matter species of a theory satisfy the strong and weak equivalence principles~\cite{Joyce:2016vqv}.
The dark sector in Einstein's Theory of General Relativity satisfies both the weak and strong equivalence principles whereas interacting dark sector models break both.
Modified gravity models satisfy the weak equivalence principle as a metric can be found, typically the Jordan frame, to which all matter species couple universally and along which they freely fall whereas they break the strong equivalence principle since this circumstance depends on a body's composition.
It is not necessary, and arguably not more natural, to cast the gravitational theory in the (Jordan) frame where (some) matter species couple minimally to the metric, and physical equations and interpretations may sometimes be more practical in Einstein frame.
The same applies for the geometric transformations in Eq.~\eqref{eq:geomdegrecast}.

\subsection{Transformations and decomposition of the metric} \label{sec:transformations}

To recast the Einstein-Hilbert action and the Einstein field equations into different geometries, particularly into Minkowski spacetime, the metric decomposition will be performed employing combinations of conformal transformations and dimensional reduction.
For completeness, the details of these transformations shall be provided here.
Readers familiar with Riemannian geometry may want to skip this section.

Generally, for the conformal transformation of a $N$-dimensional metric $g_{\mu\nu}$ with the conformal factor $\Omega^2$,
\beq
    \tilde{g}_{\mu\nu}(x^{\xi}) = \Omega^2(x^{\xi}) g_{\mu\nu}(x^{\xi}) \,, \label{eq:conftrans}
\eeq
the Ricci tensor and curvature transform as
\beqa
    R_{\mu\nu} & = & \tilde{R}_{\mu\nu} + \frac{N-2}{\Omega}\tilde{\nabla}_{\mu} \tilde{\nabla}_{\nu} \Omega + \left( \frac{1}{\Omega} \tilde{\nabla}_{\xi} \tilde{\nabla}^{\xi} \Omega - \frac{N-1}{\Omega^2} \tilde{\nabla}_{\xi} \Omega \tilde{\nabla}^{\xi} \Omega \right) \tilde{g}_{\mu\nu} \,, \\
    R & = & \Omega^2 \tilde{R} + 2(N-1)\Omega\tilde{\nabla}_{\xi} \tilde{\nabla}^{\xi} \Omega - N(N-1) \tilde{\nabla}_{\xi} \Omega \tilde{\nabla}^{\xi} \Omega \,. \label{eq:Rconf}
\eeqa
The Einstein tensor $G_{\mu\nu} = R_{\mu\nu} - \frac{1}{2}R g_{\mu\nu} \ (+ \; \Lambda g_{\mu\nu})$ transforms accordingly.

For the dimensional reduction, consider a warped spacetime metric in $N=n+m$ dimensions,
\beq
    g_{\mu\nu}(x^{\xi}) dx^{\mu}dx^{\nu} = \tilde{g}_{ij}(x^k) dx^i dx^j + F^2(x^k)\hat{g}_{AB}(x^C) dx^A dx^B \,, \label{eq:reduction}
\eeq
where Greek indices run over $n+m$ dimensions, lowercase Roman letters over $m = \dim(\tilde{M})$ dimensions and capital Roman letters over $n = \dim(\hat{M})$ co-dimensions.
The Ricci tensor and curvature can then be decomposed as
\beqa
    & & R_{ij} = \tilde{R}_{ij} - \frac{n}{F} \tilde{\nabla}_i \tilde{\nabla}_j F \,, \quad R_{AB} = \hat{R}_{AB} - \hat{g}_{AB} \left[ (n - 1) \tilde{\nabla}_k F + F \tilde{\nabla}_k \right]\tilde{\nabla}^k F \,, \label{eq:dimred1} \\
    & & R_{Ai} = 0 \,, \quad R = \tilde{R} + \frac{1}{F^2}\left[ \hat{R} - n(n-1) \tilde{\nabla}_k F - 2 n F \tilde{\nabla}_k \right] \tilde{\nabla}^k F \,. \label{eq:dimred2}
\eeqa

The conformal transformations~\eqref{eq:conftrans} and \eqref{eq:Rconf} will be applied to the Einstein-Hilbert action in Sec.~\ref{sec:fieldeqs}.
The application of the dimensional reduction of $g_{\mu\nu}$ and $R$ to the action can be conducted analogously to Ref.~\cite{Jana:2020vov} and shall not be performed explicitly here (see, however, Secs.~\ref{sec:curvature}--\ref{sec:gws} for an implementation in the matter sector).

\subsection{Conformally reframed Einstein field equations} \label{sec:fieldeqs}

For a concrete example of how the Einstein-Hilbert action and the Einstein field equations transform under the metric modifications of Sec.~\ref{sec:transformations}, consider the general conformal transformation $\tilde{g}_{\mu\nu} = \Omega^2(\tau,\mathbf{x})g_{\mu\nu}$ of the $(3+1)$-dimensional metric $g_{\mu\nu}(\tau,\mathbf{x})$.
The Einstein-Hilbert action for $g_{\mu\nu}$ reads
\beq
    S = \frac{1}{2\kappa^2} \int d^4x \sqrt{-g} \left( R - 2\Lambda \right) + \int d^4x \sqrt{-g} \mathcal{L}_m(g_{\mu\nu},\Psi_m) \,, \label{eq:action}
\eeq
where $\mathcal{L}_m$ is the Lagrangian density of the matter fields $\Psi_m$.
Under the conformal transformation this becomes
\beq
    S = \frac{1}{2\kappa^2} \int d^4x \sqrt{-\tilde{g}} \left( \Omega^{-2} \tilde{R} + 6 \Omega^{-4} \tilde{\nabla}_{\mu}\Omega \tilde{\nabla}^{\mu}\Omega - 2\Omega^{-4}\Lambda \right) + S_m[\Omega^{-2}\tilde{g}_{\mu\nu},\Psi_m] \,,
\eeq
where the matter sector is given by
\beq
    S_m[\Omega^{-2}\tilde{g}_{\mu\nu},\Psi_m] = \int d^4x \sqrt{-\tilde{g}} \Omega^{-4} \mathcal{L}_m(\Omega^{-2}\tilde{g}_{\mu\nu},\Psi_m) \,.
\eeq
Note that the matter fields are no longer minimally coupled to the metric (see Sec.~\ref{sec:degeneracy}).

For simplification, one may define $\psi \equiv \Omega^{-1}$ to arrive at the action
\beq
   \hat{S} \equiv -\frac{\kappa^2 S}{6} = \int d\tilde{V} \left[ -\frac{1}{2}\tilde{\nabla}_{\mu}\psi\tilde{\nabla}^{\mu}\psi -V(\psi) \right] - \frac{\kappa^2}{6} S_m[\psi^2\tilde{g}_{\mu\nu},\Psi_m] \label{eq:psiaction}
\eeq
with volume element $d\tilde{V}=d^4x\sqrt{-\tilde{g}}$ and potential
\beq
    V(\psi) = \frac{1}{12} \left( \tilde{R} \psi^2 - 2\Lambda \psi^4 \right) \,.
\eeq

Variation of Eq.~\eqref{eq:psiaction} with respect to the metric $\tilde{g}_{\mu\nu}$ yields the Einstein field equations
\beq
    \psi^2 (\tilde{R}_{\mu\nu} - \frac{1}{2}\tilde{g}_{\mu\nu}\tilde{R}) + 4 \left( \partial_{\mu}\psi \partial_{\nu}\psi - \frac{1}{4} \tilde{g}_{\mu\nu} \partial_{\alpha}\psi \partial^{\alpha}\psi \right) - 2\psi \left( \tilde{\nabla}_{\mu}\partial_{\nu}\psi - \tilde{g}_{\mu\nu} \tilde{\Box} \psi \right) + \tilde{g}_{\mu\nu}\Lambda\psi^4 = \kappa^2 \tilde{T}_{\mu\nu} \,, \label{eq:einstein}
\eeq
where $\tilde{T}^{\mu\nu} \equiv (2/\sqrt{-\tilde{g}}) \delta L_m/\delta\tilde{g}_{\mu\nu}$ is the energy-momentum tensor in the transformed frame with $S_m=\int d^4x L_m$ and $\tilde{\Box} = (-\tilde{g})^{-1/2} \partial_{\mu} \left( \sqrt{-\tilde{g}} \tilde{g}^{\mu\nu} \partial_{\nu} \right)$.
$\tilde{T}_{\mu\nu}$ relates to its analogous expression $T_{\mu\nu}$ from Eq.~\eqref{eq:action} as $\tilde{T}_{\mu\nu} = \psi^2 T_{\mu\nu}$ and its trace transforms as $\tilde{T} = \psi^4 T$.

The action~\eqref{eq:psiaction} and field equations~\eqref{eq:einstein} resemble those of a scalar-tensor modification of gravity, where conformal transformations are used to convert between the Einstein and Jordan frames.
Note, however, that for scalar-tensor theories, these frames interchange minimal and non-minimal couplings to the tensor field between the metric and matter sectors whereas the conformal transformation from Eq.~\eqref{eq:action} to \eqref{eq:psiaction} introduces a non-minimal coupling in both sectors, initially not coupled to $\psi$.
We shall thus refrain here from using the terminology of Einstein and Jordan frames to avoid confusion.

Varying the action~\eqref{eq:psiaction} with respect to $\psi$ yields the field equation
\beq
    \tilde{\Box}\psi - V'(\psi) = \frac{\kappa^2}{6} \psi^{-1} \tilde{T} \,. \label{eq:scal}
\eeq
Finally, note that the energy-momentum tensor is not conserved in the transformed frame,
\beq
    \tilde{\nabla}_{\mu} \tilde{T}^{\mu}_{\,\,\nu} = \psi^{-1} \tilde{T} \partial_{\nu} \psi \,. \label{eq:energy}
\eeq

\section{Casting cosmology into Minkowski space} \label{sec:minkowskicosmology}

Our Universe is well described by a perturbed expanding spatially homogeneous and isotropic spacetime metric.
The metric hereby is the solution of the Einstein field equations.
Just as with any other differential equation, one can always make a substitution of the variable, in this case the metric, one solves the equation for.
This is merely a mathematical manipulation and does not make any changes of the underlying physics (see Sec.~\ref{sec:gravity}).
But by performing transformations of the metric, one can ask whether a more useful metric, or variable, can be found by substitution that simplifies the equations or allows for some insightful physical interpretation.
A special place among the spacetimes one could transform the cosmic geometry into is occupied by the Minkowski metric, being static and flat as well as the spacetime of special relativity and quantum field theory or indeed the entire Standard Model.
It is thus worthwhile studying cosmology in the Minkowski frame.
With the metric transformations discussed in Sec.~\ref{sec:gravity}, we shall hence proceed with casting cosmology into Minkowski space (see Sec.~\ref{sec:backgrounds} for other choices).

In Sec.~\ref{sec:static}, the expanding spatially homogeneous and isotropic spacetime of the cosmic background is first cast into a static spacetime by conformal transformation.
The Friedmann equations, the evolution of energy densities, and redshift are rederived in this static frame.
Sec.~\ref{sec:curvature} then employs the combination of conformal transformations and dimensional reduction of the metric to further cast the static spacetime into Minkowski space in the presence of spatial curvature.
The transformations are then generalised to cast inhomogeneities from cosmic structure into Minkowski spacetime in Sec.~\ref{sec:curvature}.
Finally, an extension of this procedure to gravitational waves is briefly discussed in Sec.~\ref{sec:gws}.

\subsection{Cosmology in static spacetime} \label{sec:static}

The $(3+1)$-dimensional spatially homogeneous and isotropic expanding cosmological background is described by the Friedmann-Lema\^itre-Robertson-Walker (FLRW) line element
\beq
    ds^2 = -c^2 dt^2 + a(t)^2 d\mathbf{x}^2 \,, \label{eq:FLRW}
\eeq
where $t$ denotes cosmic time, $a(t)$ is the scale factor and $d\mathbf{x}^2 = (1-kr^2)^{-1}dr^2 + r^2(d\theta^2 + \sin^2\theta \: d\phi^2)$ with spatial curvature $k$.
Cast in conformal time $\tau$, Eq.~\eqref{eq:FLRW} becomes
\beq
    ds^2 = a^2(\tau)\left[-c^2 d\tau^2 + d\mathbf{x}^2 \right] \,, \label{eq:FLRW_conformal}
\eeq
where $d\tau = a^{-1}dt$.
Hence, $a^2(\tau)$ acts as a conformal factor in Eq.~\eqref{eq:FLRW_conformal}.

For simplicity briefly assuming the absence of spatial curvature ($k=0$), the metric $g_{\mu\nu}$ for $ds^2 = g_{\mu\nu}dx^{\mu}dx^{\nu}$ in Eq.~\eqref{eq:FLRW_conformal} becomes
\beq
    g_{\mu\nu} = a^2(\tau) \eta_{\mu\nu} \,,
\eeq
where $\eta_{\mu\nu} = {\rm diag}(-1, 1, 1, 1)$ is the Minkowski metric.
Thus, $g_{\mu\nu}$ is a conformal metric transformation of Minkowski spacetime with conformal factor $a^2(\tau)$.
The inverse conformal transformation $\eta_{\mu\nu} = a^{-2}(\tau) g_{\mu\nu}$ can then be used to cast the Einstein-Hilbert action into Minkowski space.
One may set $\tilde{g}_{\mu\nu}=\eta_{\mu\nu}$, $\psi = a$, and $\tilde{R} = 0$ in Eq.~\eqref{eq:psiaction} to arrive at
\beq
   \hat{S} = \int d^4x \left[ -\frac{1}{2}\partial_{\mu}\psi\partial^{\mu}\psi - V(\psi) \right] - \frac{\kappa^2}{6} S_m[\psi^2\eta_{\mu\nu},\Psi_m] \label{eq:scalefactoraction}
\eeq
with $V(\psi)=-\Lambda\psi^4/6$.

It is worthwhile noting that action~\eqref{eq:scalefactoraction} also holds for $k\neq0$, in which case, however, $\psi$ becomes both time and space dependent.
This is due to the FLRW metric generally being conformally flat with vanishing Weyl tensor, and so $g_{\mu\nu} = a^2(T,R) \eta_{\mu\nu}$ with $k\neq0$ for suitable coordinate choices $T$ and $R$~\cite{Endean:1997,Iihoshi:2007uz,Ibison:2007dv,Gron:2011yi}.
Sec.~\ref{sec:curvature} will present a different approach to recovering the gravitational action and field equations in Minkowski space for scenarios with $k\neq0$.

Consider now the conformal transformation~\eqref{eq:conftrans} of the metric $g_{\mu\nu}$ defined by the line element~\eqref{eq:FLRW}, allowing $k\neq0$, with $\psi = \Omega^{-1} = a(\tau)$ such that $\tilde{g}_{\mu\nu}$ is static.
Importantly, as highlighted in Sec.~\ref{sec:degeneracy}, given that this is merely a mathematical transformation, observed physical relations should remain unaffected by the transformation.
Indeed, as shown in the following, despite the static cosmological spacetime, the Friedmann equations and the evolution of energy densities (Sec.~\ref{sec:friedmann}) as well as redshift of distant emissions (Sec.~\ref{sec:redshift}) are recovered.

\subsubsection{Friedmann equations and energy density} \label{sec:friedmann}

Using Eqs.~\eqref{eq:conftrans} and \eqref{eq:FLRW} with $\psi = \Omega^{-1} = a(\tau)$, one finds $\tilde{R}_{00} = 0$, $\tilde{R}_{ij}= 2k\tilde{g}_{ij}$ and $\tilde{R}=6k$ from Eqs.~\eqref{eq:dimred1} and \eqref{eq:dimred2}.
Hence, for the $00$ component in Eq.~\eqref{eq:einstein}, one obtains
\beqa
    H^2 + \frac{k c^2}{a^2} - \frac{\Lambda c^2}{3} & = & -\frac{\kappa^2c^2}{3} a^{-4} \tilde{T}^0_{\,\,0} = \frac{8\pi G}{3} a^{-4} \tilde{\rho} \\
    & = & - \frac{\kappa^2c^2}{3} T^0_{\,\,0} = \frac{8\pi G}{3} \rho \,, \label{eq:00}
\eeqa
where $H \equiv \dot{a}/a = (da/dt)/a = (da/d\tau)/a^2$ denotes the Hubble function and dots indicate derivatives with respect to cosmic time $t$ throughout the article.
The $ii$ component and Eq.~\eqref{eq:00} imply
\beqa
    \frac{\ddot{a}}{a} & = & -\frac{\kappa^2c^2}{6} a^{-4} \left( -\tilde{T}^0_{\,\,0} + 3\tilde{T}^i_{\,\,i} \right) + \frac{\Lambda c^2}{3} = -\frac{4\pi G}{3} a^{-4} \left( \tilde{\rho} + \frac{3\tilde{p}}{c^2} \right) + \frac{\Lambda c^2}{3} \\
    & = & -\frac{\kappa^2c^2}{6} \left( -T^0_{\,\,0} + 3T^i_{\,\,i} \right) + \frac{\Lambda c^2}{3} = -\frac{4\pi G}{3} \left( \rho + \frac{3p}{c^2} \right) + \frac{\Lambda c^2}{3} \,. \label{eq:ii}
\eeqa
Hence, one recovers the first and second Friedmann equations.

Furthermore, Eq.~\eqref{eq:scal} yields
\beqa
    \frac{\ddot{a}}{a} + H^2 + \frac{k c^2}{a^2} - \frac{2}{3}\Lambda c^2 & = & -\frac{\kappa^2 c^2}{6} a^{-4} \tilde{T} = \frac{8\pi G}{6} a^{-4} \left( \tilde{\rho} - \frac{3\tilde{p}}{c^2} \right) \\
    & = & -\frac{\kappa^2 c^2}{6} T = \frac{8\pi G}{6} \left( \rho - \frac{3p}{c^2} \right) \,,
\eeqa
which may also be obtained from combining Eqs.~\eqref{eq:00} and \eqref{eq:ii}.
Finally, Eq.~\eqref{eq:energy} yields
\beqa
    \dot{\tilde{\rho}} & = & H (1-3w) \tilde{\rho} \,, \label{eq:rhotrans} \\
    \dot{\rho} & = & - 3 H (1+w) \rho \label{eq:rho}
\eeqa
with equation of state $w=\rho c^2/p$.
Eqs.~\eqref{eq:rhotrans} and \eqref{eq:rho} provide evolution equations for the energy densities that are consistent between both frames such that the Friedmann equations and the evolution of the scale factor with cosmic time $t$ remain invariant under the transformation.

Note that in the frame of $\tilde{g}_{\mu\nu}$, the radiation energy density $\tilde{\rho}_r$ is constant whereas the energy density associated with the cosmological constant $\tilde{\rho}_{\Lambda}$ scales as $a^4$ and that of matter $\tilde{\rho}_m$ as $a$.
Importantly, the relation of particle masses to the CMB radiation temperature remains invariant between the two frames due to the evolving masses in the Minkowski frame (see Sec.~\ref{sec:redshift}), and particles become non-relativistic at the same time.
Likewise, the relation of emitted photon energies to temperature remains invariant.
Finally, it is also worthwhile noting that the evolution of $\psi=a$ and the evolution of the energy densities can be recovered without the Einstein field equations and thus the metric variation of action~\eqref{eq:psiaction}, only using Eqs.~\eqref{eq:scal} and \eqref{eq:energy}.

\subsubsection{Redshift} \label{sec:redshift}

Importantly, as shown in the following, despite the static geometry of the Universe in Sec.~\ref{sec:friedmann}, cosmological observations are still redshifted.
However, rather than due to the expanding space, redshift is caused here by evolving particle masses~\cite{Domenech:2016yxd, Dalang:2019fma} (also see Ref.~\cite{Wetterich:2013aca}).
Thus, a photon emitted in the past as the result of some atomic line transition will be redshifted with respect to the measured emission of the same transition in our local, present laboratories.
Photons still follow null geodesics in the conformally transformed frame~\cite{Dalang:2019fma} and our detectors measure them at their emission frequencies.

To illustrate the evolution of masses, consider a matter field with the simple Lagrangian
\beq
    L_m = \sqrt{-g}\left[ -\frac{1}{2}g^{\mu\nu}\partial_{\mu}\Psi_m \partial_{\nu}\Psi_m - \frac{1}{2} \left(\frac{mc}{\hbar}\right)^2 \Psi_m^2 \right] \label{eq:matter} \,,
\eeq
which yields the field equation $\Box\Psi_m - (mc/\hbar)^2 \Psi_m = 0$.
After conformal transformation, this becomes
\beq
    L_m = \sqrt{-\tilde{g}} \psi^2 \left[ -\frac{1}{2}\tilde{g}^{\mu\nu}\partial_{\mu}\Psi_m \partial_{\nu}\Psi_m - \frac{1}{2} \left(\frac{mc\psi}{\hbar}\right)^2 \Psi_m^2 \right] \label{eq:transformedmatterlagrangian}
\eeq
and
\beq
    \tilde{\Box}\Psi_m - (mc/\hbar)^2\psi^2 \Psi_m = 0 \,, \label{eq:evolvingmass}
\eeq
where one may define the evolving mass $\tilde{m} \equiv m \psi$ in this frame.
This evolution of $\tilde{m}$ is also more generally found for spinor fields interacting with a vector field in curved spacetime~\cite{Domenech:2016yxd, Dalang:2019fma} and has important implications for our observations.

Consider for example the transition lines from an $n_i$ state to an $n_f$ state of a hydrogen atom, for which one finds
\beq
    \tilde{E}_{n_i n_f} = \frac{\alpha^2 \tilde{m}_e}{4\pi\hbar}\left( \frac{1}{n_i^2} - \frac{1}{n_f^2} \right) = \psi E_{n_i n_f} \,.
\eeq
The redshift $\tilde{z}$ is given by the ratio between the energy emitted from this transition as observed in the laboratory $\tilde{E}_{n_i n_f}$ and the frequency observed from this transition occurring at a distant galaxy $\tilde{\omega}_o$.
One finds
\beq
    1+\tilde{z} = \frac{\tilde{E}_{n_i n_f}(O)}{\hbar\tilde{\omega}_o} = \frac{\psi_o}{\psi_{s}} \frac{\tilde{E}_{n_i n_f}(S)}{\hbar\tilde{\omega}_o} = \frac{\psi_o}{\psi_{s}} \frac{\tilde{E}_{n_i n_f}(S)}{\hbar\tilde{\omega}_s} = \frac{\psi_o}{\psi_{s}} \,, \label{eq:redshift1}
\eeq
using that the photon frequency does not change with its propagation through Minkowski space, $\tilde{\omega}_o = \tilde{\omega}_s$.
Inversely, one has
\beq
    \frac{\psi_o}{\psi_{s}} = \frac{\psi_o}{\psi_{s}} \frac{\tilde{E}_{n_i n_f}(S)}{\hbar\tilde{\omega}_s} = \frac{\psi_o}{\psi_{s}} \frac{\tilde{E}_{n_i n_f}(S)}{\hbar\tilde{\omega}_o} = \frac{E_{n_i n_f}(S)}{\hbar\omega_o} = \frac{\omega_s}{\omega_o} = (1+z) \,. \label{eq:redshift2}
\eeq
This implies that redshift is invariant between the two frames, as expected for an observable quantity, and is determined by the ratio between the generally spacetime dependent $\psi(\tau,\mathbf{x})$ evaluated at source and observer.

Specifically, for the spatially homogeneous and isotropic case, $\psi_s = a(t_{em}) = a$ and $\psi_o = a(t_0) = a_0$, and thus
\beq
    \frac{\psi_o}{\psi_s} = \frac{a_0}{a} = 1+z \,.
\eeq

Note that redshift is also recovered for gravitational wave emissions, which is also attributed to the evolution of masses~\cite{Dalang:2019fma}, where massless gravitons, like photons, still follow null geodesics in the conformally transformed frame.

\subsection{Spatial curvature} \label{sec:curvature}

As pointed out in Sec.~\ref{sec:static}, spatially curved FLRW metrics have vanishing Weyl tensor and are thus conformally flat.
They can therefore be written as $g_{\mu\nu} = a^2(T,R) \eta_{\mu\nu}$ for suitable coordinate choices $T$ and $R$~\cite{Endean:1997,Iihoshi:2007uz,Ibison:2007dv,Gron:2011yi}.
In principle, one could thus directly apply the conformally reframed Einstein field equations of Sec.~\ref{sec:fieldeqs}.
Another procedure to transform the gravitational action and field equations into Minkowski space is to perform a combination of dimensional reduction and conformal transformations.

Specifically, the static line element obtained after the conformal transformation in Sec.~\ref{sec:friedmann} is given by
\beqa
    d\tilde{s}^2 = \tilde{g}_{\mu\nu} dx^{\mu}dx^{\nu} & = & -c^2 d\tau^2 + \frac{dr^2}{1-kr^2} + r^2(d\theta^2 + \sin^2\theta \: d\phi^2) \label{eq:curvedFLRW} \\
    & = & -c^2 d\tau^2 + d\chi^2 + r^2(\chi) (d\theta^2 + \sin^2\theta \: d\phi^2) \\
    & = & -c^2 d\tau^2 + \left[\frac{r(\zeta)}{\zeta}\right]^2 \left[d\zeta^2 + \zeta^2(d\theta^2 + \sin^2\theta) \: d\phi^2 \right] \label{eq:curvedFLRWtransformed}
    \,,
\eeqa
where hyperspherical coordinates are adopted in the second line and the coordinate change $d\chi = \left[r(\zeta)/\zeta\right] d\zeta$ is performed in the third.
Note that for $k=0$, $r(\chi) = \chi$ and hence $d\zeta=d\chi$.
To now cast the Einstein-Hilbert action~\eqref{eq:psiaction} for $\tilde{g}_{\mu\nu}$ and $\psi = a$ with $k\neq0$ into Minkowski space, one can apply Eq.~\eqref{eq:reduction} to perform the dimensional reduction of $\tilde{R}$ to separate out the spatial Ricci scalar $\tilde{R}^{(3)}$ in the action and subsequently perform a conformal transformation of $\tilde{R}^{(3)}$ with $\hat{g}^{(3)}_{ij} = [r(\zeta)/\zeta]^{-2} \tilde{g}^{(3)}_{ij} = \eta^{(3)}_{ij}$.

The procedure shall not be carried out for the metric sector here.
Of interest is rather the transformation of the matter sector.
Revisiting the toy matter Lagrangian~\eqref{eq:transformedmatterlagrangian}, one finds in Minkowski space with Cartesian coordinates the scalar field equation
\beqa
 -\frac{1}{c^2}\partial_{\tau}^2\Psi_m + \frac{1}{\xi^2} \sum_i \left( 1 + \frac{(\ln \xi)_{,i}}{(\ln \partial_i \Psi_m)_{,i}} \right) \partial_i^2 \Psi_m - \frac{m^2c^2\psi^2}{\hbar}\Psi_m & = & 0 \,, \\
 -\frac{1}{c^2}\partial_{\tau}^2\Psi_m + \frac{\ell(x^{\mu})^2}{\ell_p^2} \sum_i \partial_i^2 \Psi_m - \frac{m^2c^2\psi^2}{\hbar}\Psi_m & = & 0 \,, \label{eq:confcurved}
\eeqa
where $\xi \equiv r(\zeta)/\zeta$.
Spatial curvature may thus be interpreted as a varying length scale $\ell(x^{\mu}) = \ell(\xi, \xi_{,i}, \Psi_{m,i}, \Psi_{m,ii})$ in Minkowski space.
Note that for flat space $k=0$, $r=\zeta$ and $\xi=1$, so that $\ell=\ell_p$.

\subsection{Cosmic structure} \label{sec:structure}

Our Universe is of course not perfectly spatially homogeneous and isotropic, and cosmological metric perturbations are typically not conformal (see Secs.~\ref{sec:nonlinearities} and \ref{sec:tensions} for conformal perturbations).
But performing an analogous procedure to that of Sec.~\ref{sec:curvature}, one may also recast the Einstein-Hilbert action for the perturbed Cosmos into Minkowski space.

For that, consider the metric $g_{\mu\nu}$ defined by the line element
\beq
    ds^2 = g_{\mu\nu} dx^{\mu}dx^{\nu} = a^2(\tau)\left[-c^2 e^{2\alpha(\tau,\mathbf{x})} d\tau^2 + e^{2\beta(\tau,\mathbf{x})} d\mathbf{x}^2 \right] \,, \label{eq:metric}
\eeq
which for $\alpha \equiv \Psi$, $|\alpha| \ll 1$ and $\beta \equiv \Phi$, $|\beta| \ll 1$  becomes the usual linearly perturbed FLRW line element in conformal Newtonian (or longitudinal) gauge and conformal time,
\beq
    ds^2 = a^2(\tau) \left[ -c^2 (1+2\Psi(\tau,\mathbf{x})) d\tau^2 + (1+2\Phi(\tau,\mathbf{x})) d\mathbf{x}^2 \right] \,. \label{eq:linmetric}
\eeq
For the following discussion, the perturbations may not need to be linear, however.
Note that $d\mathbf{x}^2$ may not be flat ($k\neq0$), in which case a factor $r(\zeta)/\zeta$ from Eq.~\eqref{eq:confcurved} can first be absorbed into $e^\beta$.
Next, one can apply the conformal transformation $\tilde{g}_{\mu\nu}=a^{-2}(\tau)e^{-2\alpha(\tau,\mathbf{x})}g_{\mu\nu}$, then perform the dimensional reduction to $\tilde{R}^{(3)}$ and the conformal transformation of that with $\hat{g}^{(3)}_{ij} = e^{2(\alpha-\beta)} \tilde{g}^{(3)}_{ij} = \eta^{(3)}_{ij}$ such that the Einstein-Hilbert action is recast into Minkowski space.

We shall again refrain from computing the resulting metric sector for this frame and instead focus on the scalar field equation for $\Psi_m$.
The transformations yield
\beqa
 -\frac{1}{c^2}\left(1+\frac{3\gamma_{,\tau}}{(\ln \Psi_{m,\tau})_{,\tau}}\right) \partial_{\tau}^2\Psi_m + e^{-2\gamma} \sum_i \left( 1 + \frac{\gamma_{,i}}{(\ln \Psi_{m,i})_{,i}} \right) \partial_i^2 \Psi_m & & \nonumber\\
 - \left(\frac{ae^{\alpha}mc}{\hbar} \right)^2 \Psi_m & = & 0 \,, \label{eq:prescales} \\
 -\frac{1}{c^2}\frac{\iota^2(x^{\mu})}{\iota_0^2}\partial_{\tau}^2\Psi_m + \frac{\ell^2(x^{\mu})}{\ell_0^2} \sum_i \partial_i^2 \Psi_m - \frac{m^2(x^{\mu})c^2}{\hbar^2} \Psi_m & = & 0  \label{eq:scales}
\eeqa
in Cartesian coordinates, where $\gamma \equiv \beta-\alpha$.
Hence, one encounters a variation of time $\iota$, length $\ell$, and mass scales $m$ for the field $\Psi_m$ in Minkowski space.
Note that one could also recast Eq.~\eqref{eq:scales} as a varying speed of light $c(x^{\mu})$ and/or a varying Planck constant $\hbar(x^{\mu})$.
Similarly, one would interpret the variation of $\ell^2(x^{\mu})c^3(x^{\mu})/\hbar(x^{\mu}) = G(x^{\mu})$ as a variation of Newton's gravitational constant.

Importantly, the same metric transformations performed on the metric part of the action (and on the other matter fields) will leave physics invariant.
Predictions for observable quantities will remain the same, but interpretations may change such as for the redshift in Sec.~\ref{sec:redshift}.
For a loose analogy, consider the observed angular positions of planets in the sky, whose predicted motions are unaffected by whether we adopt a heliocentric view or an exact transformation of that into geocentric coordinates.
The heliocentric system provides a much simpler and universal physical picture than the resulting epicycles, but one would not adopt heliocentric coordinates to describe a small-scale experiment in a laboratory on Earth.
Similarly, the FLRW metric may not be optimal for describing small-scale physical processes in our nonlinear Universe, and a change of mathematical frame may become more suitable (see Sec.~\ref{sec:applications}).

It is worthwhile noting that if not performing the analogous transformations in the metric sector, or only performing them in the metric but not in the matter sector, one does modify physics.
For instance, if not applying the conformal transformation for the matter sector in Eq.~\eqref{eq:psiaction}, one recovers a scalar-tensor modification of gravity.
One may furthermore consider non-universal couplings of different matter species to this conformal factor, thereby breaking the weak equivalence principle~\cite{Joyce:2016vqv} (see Sec.~\ref{sec:degeneracy}).
This multitude of possibilities motivates generally testing, across the vast spacetime, for phenomenological variations in the effective time, length and mass scales of the observed behaviour of different matter species as in Eq.~\eqref{eq:scales}, or likewise in the fundamental couplings.
Note in this context that while the Minkowskian picture is physically equivalent to the standard cosmological picture both at the classical and quantum level~\cite{Postma:2014vaa} and observationally indistinguishable from it, it offers a different basis for theoretical extensions.
These may appear more natural in one picture over the other and if verified, could favour one framework over the other.

\subsection{Gravitational waves} \label{sec:gws}

Finally, consider the symmetric traceless tensor perturbation $h_{ij}$ of $d\mathbf{x}^2$ in Eq.~\eqref{eq:metric}.
This term will remain in $\hat{g}^{(3)}_{ij}$ after the transformations of Secs.~\ref{sec:curvature} and \ref{sec:structure}.
To then cast the Einstein-Hilbert action into Minkowski space, one may perform a metric transformation of the form $\hat{g}^{(3)}_{ij} = A_{ij}^{\:\: kl} \eta_{kl}^{(3)} + B_{ij}$.
With suitable transformations and coordinates, this amounts to simply separating out the tensor field $h_{ij}$ in addition to the scalar fields from the three and four-dimensional conformal transformations.
One can see from Eq.~\eqref{eq:scales} that this will lead to oscillations in time and length scales.
Since gravitational waves are not studied further in this article, a more detailed calculation is left to future work.

\section{Casting cosmology into other backgrounds} \label{sec:backgrounds}

The transformations of Sec.~\ref{sec:gravity} have been used in Sec.~\ref{sec:minkowskicosmology} to cast cosmology into static spacetime and then further into Minkowski space.
The target cosmological background may however not need to be Minkowski space.
For illustration, in Sec.~\ref{sec:EdS}, our Universe is cast into Einstein-de Sitter spacetime, which will provide room for a reinterpretation of dark energy.
In Sec.~\ref{sec:dS}, it is cast into a de Sitter background, which will allow for a reinterpretation of dark matter.
Sec.~\ref{sec:nonlinearities} then casts nonlinear inhomogeneities of the metric into a smooth spacetime with implications for the Hubble tension.

For simplicity, in the following, radiation will be neglected and a vanishing spatial curvature is assumed ($k=0$), but results can be generalised for the presence of those.
For a $\Lambda$CDM cosmology, one has
\beqa
    H^2 & = & t_{\Lambda}^{-2} \left( 1 + \left( \frac{a}{a_{eq}} \right)^{-3} \right) \,, \label{eq:hubble} \\
    a & = & a_{eq} \sinh^{2/3}\left( \frac{3}{2} \frac{t}{t_{\Lambda}} \right) \,,
\eeqa
where $a_{eq}$ denotes the scale factor at equality between the energy densities in matter and in the cosmological constant, and $t_{\Lambda}$ is the time scale associated with the cosmological constant.
Specifically,
\beq
    a_{eq} = a_0 \left( \frac{\Omega_m}{1-\Omega_m} \right)^{1/3} \,, \quad \quad t_{\Lambda} = \left( \frac{\Lambda c^2}{3} \right)^{-1/2} \,,
\eeq
where the matter density parameter is defined today $t_0$ with $\Omega_m \equiv \left(\kappa^2\rho_m / (3 H^2) \right)(t=t_0)$.
Of interest here will be the two limits of $\Lambda$ and matter domination,
\beqa
    a \gg a_{eq} & : & \left\{ \begin{array}{rcl} H & = & t_{\Lambda}^{-1} \\ a & = & 2^{-2/3} a_{eq} e^{t/t_{\Lambda}} = t_{\Lambda} (\tau_{\infty} - \tau)^{-1} \\ \tau & = & \tau_{\infty} - 2^{2/3}a_{eq}^{-1} t_{\Lambda} e^{-t/t_{\Lambda}} \end{array} \right. \\
    & & \\
    a \ll a_{eq} & : & \left\{ \begin{array}{rcl} H & = & t_{\Lambda}^{-1}(a/a_{eq})^{-3/2} \\ a & = & a_{eq} \left(\frac{3}{2} \frac{t}{t_{\Lambda}}\right)^{2/3} = \frac{a_{eq}^3}{4t_{\Lambda}^2} \tau^2 \\ \tau & = & \frac{3}{a_{eq}} \left(\frac{2t_{\Lambda}}{3}\right)^{2/3} t^{1/3} \end{array} \right.
\eeqa
for the late-time de Sitter and early-time Einstein-de Sitter eras, respectively.

Rather than transforming the FLRW metric with scale factor $a_{{\Lambda}\rm CDM}$ into static spacetime, the conformal factor $\psi^{-2}$ is now used to cast the Universe into another FLRW metric with different expansion $\tilde{a}_{\rm target}$.
Thus, $\psi = a_{{\Lambda}\rm CDM}/\tilde{a}_{\rm target}$ such that
\beq
    \tilde{g}_{\mu\nu} = \psi^{-2} g_{\mu\nu} = \left( \frac{\tilde{a}_{\rm target}}{a_{{\Lambda}\rm CDM} } \right)^2 a_{{\Lambda}\rm CDM}^2 \eta_{\mu\nu} = \tilde{a}_{\rm target}^2 \eta_{\mu\nu} \,.
\eeq
Importantly, it follows from Eq.~\eqref{eq:evolvingmass} that the mass of the matter field $\Psi_m$ is now evolving as
\beq
    \tilde{m} = m\psi = m \frac{a_{{\Lambda}\rm CDM}}{\tilde{a}_{\rm target}} \,.
\eeq
Furthermore, while conformal time $\tau$ remains the same, cosmic time changes due to the conformal transformation, $d\tilde{t} = (\tilde{a}_{\rm target}/a_{{\Lambda}\rm CDM})dt$.

For convenience, in the following, the indices `$\Lambda$CDM' and `target' will be dropped.

\subsection{Einstein-de Sitter background} \label{sec:EdS}

In an Einstein-de Sitter universe, the scale factor evolves as $\tilde{a} \propto \tilde{t}^{2/3} \propto \tau^2$.
Since $\Lambda$CDM is Einstein-de Sitter at early times, in the limit of $\tau\rightarrow0$, one has $\tilde{a}=a$ and $\tilde{t}=t$.
The scale factor can be normalised at $a_{eq}$ as in Eq.~\eqref{eq:hubble}.
Given the absence of an era of $\Lambda$ domination, in Einstein-de Sitter spacetime, $a_{eq}$ can be thought of as a normalisation for when $c^2\kappa^2\tilde{\rho}$ falls below a given threshold $\Lambda$.
It follows that at early times, $\tilde{m} = m$, and at late times,
\beq
    \tilde{m} = \left(\frac{t_{\Lambda}}{a_{eq}}\right)^3\frac{4}{\tau^2(\tau_{\infty}-\tau)} m \,,
\eeq
approaching infinity in finite conformal time $\tau_{\infty}$.

Hence, cosmic acceleration can be interpreted in this frame as increasing and diverging particle masses that start to evolve once cosmic time approaches $\tilde{t} \sim t_{\Lambda}$ set by the threshold $\Lambda$.

\subsection{De Sitter background} \label{sec:dS}

In de Sitter spacetime, one has a constant Hubble function $\tilde{H} = t_{\Lambda}^{-1}$, the scale factor evolves as $\tilde{a} = \tilde{a}_{dS} e^{\tilde{t}/t_{\Lambda}} = t_{\Lambda}(\tilde{\tau}_{\infty}-\tau)^{-1}$, and conformal and cosmic times relate as $\tau = \tilde{\tau}_{\infty} - \tilde{a}_{dS}^{-1} t_{\Lambda} e^{-\tilde{t}/t_{\Lambda}}$.
Normalising $\tilde{\tau}_{\infty} = \tau_{\infty}$ by suitable choice of $\psi$, one finds that $\tilde{m} = m$ at late times and
\beq
    \tilde{m} = \frac{1}{4} \left(\frac{a_{eq}}{t_{\Lambda}}\right)^3 \tau^2(\tau_{\infty}-\tau) m \simeq \frac{\tau_{\infty}}{4} \left(\frac{a_{eq}}{t_{\Lambda}}\right)^3 \tau^2 m
\eeq
at early times.

Hence, similarly to interpreting dark energy as the late-time divergent increase of particle masses, in this frame, one may interpret dark matter as the effect of an early-time increase of particle masses with $\tau^2$ that converges to a constant mass $m$ at late times.

\subsection{Nonlinear inhomogeneities} \label{sec:nonlinearities}

Rather than casting the cosmic structure encompassed in the nonlinearly perturbed FLRW metric~\eqref{eq:metric} into Minkowski space, one may cast it into smooth FLRW spacetime instead.
This will lead to a similar interpretation of changing mass, length and time scales as in Sec.~\ref{sec:structure}, which vary locally with respect to the cosmological background.

The following discussion will focus on conformal inhomogeneities, which are a special type of nonlinear inhomogeneities that have not been studied in much detail.
Particularly, in Refs.~\cite{Lombriser:2019ahl, Bose:2020cjb}, they were proposed as a simultaneous solution of cosmological tensions in the measured Hubble constant $H_0$, the amplitude of matter fluctuations $\sigma_8$, and the amplitude of CMB lensing $A_L$, which would be found from residing in a local conformal void.
The nonlinearly conformally perturbed FLRW metric has also been dubbed Conformal Friedmann-Lema\^itre-Robertson-Walker (CFLRW) cosmology in Ref.~\cite{Visser:2015iua}.

Consider the scenario where we occupy a local conformal inhomogeneity in a $\Lambda$CDM FLRW (or CFLRW) universe such that $a_{\Lambda \rm CDM}(\tau) \rightarrow a_{\Lambda \rm CDM}(\tau, \mathbf{x}) \equiv C(\tau, \mathbf{x})\bar{a}_{\Lambda \rm CDM}(\tau)$ with conformal factor $C^2$.
Proceeding as in Secs.~\ref{sec:EdS} and \ref{sec:dS}, let the target cosmology be a smooth $\Lambda$CDM FLRW metric, $\tilde{a}_{\Lambda \rm CDM}(\tau) = \bar{a}_{\Lambda \rm CDM}(\tau)$.
This implies the evolution of masses in this frame as
\beq
    \tilde{m} = \frac{a_{\Lambda \rm CDM}(\tau, \mathbf{x})}{\bar{a}_{\Lambda \rm CDM}(\tau)} m = C(\tau, \mathbf{x}) m \,.
\eeq
Eqs.~\eqref{eq:redshift1} and \eqref{eq:redshift2} imply that redshifts relate as $1+z = C_0 (1 + \bar{z}) / C(\tau_s,\mathbf{x}_s) = 1+\tilde{z}$, where the conformal factor $C_0^2 \equiv C^2_{obs}$ is defined at the spacetime location of the observer and $\bar{z}$ is the redshift one would obtain if both source and observer resided in the smooth background.
Defining masses at the location of the observer, one has
\beq
    \tilde{m} = \frac{C(\tau,\mathbf{x})}{C_0} \tilde{m}_0 \,.
\eeq

For a simple application, one may adopt a top-hat inhomogeneity of radius $R$ with $C = (1+\delta)^{-1/3}$, where $\rho_m = \bar{\rho}_m (1+\delta)$.
At the spacetime location of the observer, $C_0 = (1+\delta_0)^{-1/3}$, and for a source outside the top hat in the smooth background, $C_s = 1$.
In this case, masses evolve as $\tilde{m} = m = C_0^{-1} \tilde{m}_0 = (1+\delta_0)^{1/3} \tilde{m}_0$ and redshifts relate as $1 + z = C_0 (1 + \bar{z}) = (1+\delta_0)^{-1/3} (1 + \bar{z}) = 1 + \tilde{z}$.

This implies that if assuming particles, or larger massive objects, to be located in the smooth cosmological FLRW background, one ought to correct for the local evolution of their masses and adapt redshifts accordingly to correctly interpret observations.

Let us briefly consider the consequences of failing to do so.
The matter contribution to the Hubble function in the background is given by
\beq
    \bar{H}^2  \supset \frac{\kappa^2\bar{\rho}_{m0}}{3} (1 + \bar{z} )^3 = \frac{\kappa^2\bar{\rho}_{m0}}{3} (1 + \delta_0) (1 + z )^3 \,.
\eeq
Importantly, as discussed in Sec.~\ref{sec:redshift}, one observes the frame-invariant redshift $z = \tilde{z}$ and not $\bar{z}$.
Thus, wrongly assuming $\bar{z} = z$ for a source outside the top hat with background density $\bar{\rho}_m$ will give $(1+\delta_0)^{-1/2} \bar{H}_0$ rather than the Hubble parameter $\bar{H}_0$.
Hence, if we reside in a conformal top-hat underdensity $\delta_0 < 0$ and interpret our observations by placing ourselves into the smooth cosmological background instead, we infer an enhanced effective Hubble parameter.
Specifically, a $\sim$10\% enhancement of $\bar{H}_0$ requires a $\sim$20\% conformal underdensity~\cite{Lombriser:2019ahl, Bose:2020cjb}.
Note that the size of the conformal void $R$ does not need to be large to change the inferred Hubble parameter as opposed to that of conventional voids.
A diameter of a few tens of megaparsecs may be sufficient~\cite{Lombriser:2019ahl, Bose:2020cjb}.
This scenario will be further investigated in Sec.~\ref{sec:tensions}.

\section{Applications} \label{sec:applications}

Casting cosmology into Minkowski space in Sec.~\ref{sec:minkowskicosmology} has revealed a reinterpretation of cosmic expansion as an evolution of particle masses with time (also see Ref.~\cite{Wetterich:2013aca}) and more generally, a variation of time, length and mass scales across spacetime for a perturbed and spatially curved Cosmos.
Alternatively, adopting Einstein-de Sitter and de Sitter backgrounds in Secs.~\ref{sec:EdS} and \ref{sec:dS} suggested an interpretation of dark energy and dark matter as evolving particle masses that respectively diverge and converge with proceeding age of the Universe.
As described in Secs.~\ref{sec:nonlinearities}, conformal inhomogeneities defined by a conformal factor separating them from the smooth cosmological background offer a potential solution to the Hubble tension (also see Refs.~\cite{Lombriser:2019ahl, Bose:2020cjb}).

The applications of geometric transformations of the cosmic spacetime are manifold, and the following discussions will adumbrate a few particularly promising ones.
In Sec.~\ref{sec:cosmologicalconstant}, the cosmological constant problem is revisited in the cosmological Minkowski frame of Sec.~\ref{sec:minkowskicosmology}.
Sec.~\ref{sec:tensions} further examines the conformal inhomogeneities and in particular a local conformal cosmic void as a simultaneous solution of the $H_0$, $\sigma_8$, and $A_L$ tensions and identifies further observable implications of those.
Finally, in Sec.~\ref{sec:dm}, candidates for dark matter, inflation and baryogenesis are identified that emerge from the formalism of Sec.~\ref{sec:static}.
It should be emphasised that the discussions provided here are far from exhaustive and should thus only serve as stimulus for further investigation.

\subsection{Naturalness of the cosmological constant} \label{sec:cosmologicalconstant}

In light of the importance of the cosmological constant problem (Sec.~\ref{sec:intro}), it is worthwhile revisiting it in the Minkowski frame of Sec.~\ref{sec:minkowskicosmology} for a possible reinterpretation of the problem.
Sec.~\ref{sec:scalefactorfield} examines the naturalness of the value of the cosmological constant for the Universe cast into Minkowski space.
Sec.~\ref{sec:quantisation} inspects the consistency with the gravitating ground state energy density from canonical quantisation in this frame.
In this process, the scale factor of the metric will be treated as a scalar field and quantised.
It should be emphasised, however, that one could instead proceed analogously with the quantisation of a matter field $\Psi_m$ as given by Eq.~\eqref{eq:matter} to arrive at the same conclusions.

\subsubsection{The scale factor as a scalar field} \label{sec:scalefactorfield}

For the following discussion, let us for simplicity assume $\tilde{R}=0$ and $\tilde{T}=0$ and redefine $\psi \rightarrow i \psi$ in Eq.~\eqref{eq:psiaction}.
Note that the new field $\psi$ must not need to be imaginary as one may also take $a\rightarrow ia$, representing a rotation of metric signature, such that $\psi=a$ remains real.
This gives the scalar field action
\beq
    S_{\psi} = \int d^4x \left[ - \frac{1}{2} \partial_{\mu}\psi \partial^{\mu}\psi - V(\psi) \right] \,, \label{eq:scalfieldaction}
\eeq
where $V = \Lambda \psi^4/6$.
The scalar field equation becomes
\beq
    \tilde{\Box} \psi - \frac{2}{3} \Lambda \psi^3 \,, \label{eq:rotscalfield}
\eeq
where one may identify the effective mass $m^2 = V''(\psi) = 2\Lambda (\psi\hbar/c)^2$.
More formally, consider a small perturbation $\delta \psi$ around $\psi_* = a(\tau_*)$ such that $\psi = \psi_*(1+\delta \psi)$.
Additionally making the field redefinition $\chi = 1 + 2\delta\psi$, the potential can be rewritten as $6V/\Lambda = \psi^4 \simeq \psi_*^4(1+2\delta\psi)^2 = \psi_*^4\chi^2$.
Variation of the action with respect to $\chi$ yields the field equation $\tilde{\Box}\chi - (4/3)\Lambda\psi_*^2\chi$ with the Klein-Gordan mass given by $m^2_{\chi} = (4/3)\Lambda (\psi_*\hbar/c)^2$.
The mass associated with the field $\psi$ thus varies as
\beq
    m \sim \frac{\hbar}{c} \sqrt{\Lambda} a \,. \label{eq:evolvingplanckmass}
\eeq
This time dependence of $m$ has important implications for the question of the naturalness of $\Lambda$ as inferring its value from a mass scale, $\Lambda \sim m^2/a^2$, implies a choice of $a$.

Consider the Planck time $t_p$ with $a_p \equiv a(t_p)$ and the present $t_0$ with $a_0 \equiv a(t_0)$.
Let us furthermore normalise the scale factor at Planck time, $a_p=1$ (one could also normalise masses as $\hat{m} = m/a_p$ instead), and assume that the volume of our currently observable Universe originated from a Planck volume at Planck time $\ell_u^3 = (a_0/a_p)^3\ell_p^3$ with Planck length $\ell_p$.
Our observations imply $\ell_u\sim10^{61}\ell_p$.
For a Plank mass $m_p$ of the field $\psi$ today, $m_0=m_p$, one therefore finds
\beq
    \Lambda \sim a_0^{-2} \frac{c^2}{\hbar^2} m_0^2 = a_0^{-2} \frac{c^2}{\hbar^2} m_p^2 = a_0^{-2} \ell_p^{-2} \sim \left(\frac{\ell_p}{\ell_u}\right)^2 \ell_p^{-2} \sim 10^{-122} \ell_p^{-2} \,.
\eeq
Thus, in this frame, $\Lambda$ takes on a natural value set by the Planck scale and the size of the present Universe.

This may not seem surprising given that we are located at an era of equality between $\Lambda$ and $\kappa^2c^2\rho_m$, where the size of the Universe is set by the scale of $\Lambda$, $\ell_u \sim \ell_{\Lambda} \equiv c t_{\Lambda}$.
Importantly, however, one should consider $m_*=\sqrt{\Lambda}\hbar c^{-1}$ as the fundamental mass scale whereas the field mass $m_{\chi}$ is subject to the time of evaluation of $a(t)$ in Eq.~\eqref{eq:evolvingplanckmass} and only becomes $m_p$ at $a_0$.
The fact that $\ell_u \sim \ell_{\Lambda}$ is therefore attributed to the coincidence or \emph{Why now?}~problem of why $\Lambda \kappa^{-2} \approx \rho_m c^2$ today whereas these energy densities differ by orders of magnitude in the past and future of the Universe.
To make this connection, note that $\ell_u/\ell_p \sim t_0/t_p$.
Matter domination implies $H_0\sim t_0^{-1}$ and when $\Lambda$ dominates, $H\sim t_{\Lambda}^{-1}$.
Hence, at matter-$\Lambda$ equality, $t_0 \approx t_{\Lambda}$ such that $\ell_u/\ell_p \sim t_{\Lambda}/t_p$.

A solution to the coincidence problem in turn is provided by anthropic considerations.
This can simply be in the sense of an observer's bias on the observer's location in cosmic history without the need to invoke a multiverse.
For that, assume a universe configured by the fundamental mass scale given by $\sqrt{\Lambda}\hbar c^{-1}$ and the matter energy density $\rho_m$ at their observed values.
As described in Ref.~\cite{Blanco:2020ipk}, this universe produces a star formation peak at the age of $\sim 4$~Gyr whereas terrestrial planet formation peaks at $\sim 8$~Gyr, making Earth a typical sample from this distribution.
If the time it took on Earth to develop observers can be considered typical, the most likely epoch for observers to emerge in the Universe is around the equality between $\Lambda$ and $\kappa^2c^2\rho_m$, setting $t_0 \approx t_{\Lambda}$, $\ell_u \approx \ell_{\Lambda}$ and $m_p \sim a(t_0) \sqrt{\Lambda} \hbar c^{-1}$.

Thus, neither the observed value of $\Lambda$ nor the comparable size of $\kappa^2c^2\rho_m$ today seem problematic in Minkowski space.

\subsubsection{Canonical quantisation of the scale factor} \label{sec:quantisation}

For completeness, let us compute $\Lambda$ from the ground state energy density attributed to canonical quantisation of the field $\psi$.
For the Lagrangian density $\mathcal{L}$ in Eq.~\eqref{eq:rotscalfield}, the canonical momentum and classical Hamiltonian are given by
\beq
 \pi = \frac{\delta \mathcal{L}}{\delta \partial_0\psi} \,, \quad \quad H = \int d^3x \left[ \pi \partial_0\psi - \mathcal{L} \right] \,,
\eeq
where the model is quantised with the commutator
\beq
 \left[ \hat{\psi}({\bf x}), \hat{\pi}({\bf y}) \right] = i \hbar \delta({\bf x} - {\bf y}) \,.
\eeq
As in Sec.~\ref{sec:scalefactorfield}, we shall define the Klein-Gordan field $\chi$ with mass $m_{\chi}$ for small perturbations around $\psi_*$.
The ground state energy density for that is
\beq
 \epsilon_0 = \frac{1}{2} \int \frac{d^3k}{(2\pi)^3} \sqrt{(k\hbar c)^2 +(m_{\chi}c^2)^2} \,.
\eeq
Taking $m_{\chi}$ as the ultraviolet cutoff, one obtains
\beq
    \epsilon_0 \sim \frac{m_{\chi}^4 c^5} {\hbar^3} \sim \hbar c \Lambda^2 a^4 \,.   
\eeq
For the gravitating vacuum energy density in Minkowski space, one therefore has
\beq
    \frac{\hbar}{m_*^2 c^3} \epsilon_0 \sim \frac{\hbar^2}{m_*^2 c^2} \Lambda^2 a^4 \sim \Lambda a^4 = \kappa^2 c^2 \tilde{\rho}_{\Lambda} \,.
\eeq
Since $\tilde{\rho}_{\Lambda}(a) = \rho_{\Lambda} a^4$ (see Sec.~\ref{sec:static}),
one recovers
\beq
    \kappa^2 c^2 \rho_{\Lambda} = \Lambda \,.
\eeq

\subsection{Observed cosmological tensions} \label{sec:tensions}

A conformal transformation was used in Sec.~\ref{sec:nonlinearities} to cast a universe where we occupy a conformal top-hat inhomogeneity into one where we reside in the smooth cosmological FRLW background geometry.
The implication of that was that masses of particles and larger objects as well as redshifts would rescale with the local density field.
Failing to account for these modifications shifts inferred cosmological quantities such as the cosmic expansion rate.
Refs.~\cite{Lombriser:2019ahl, Bose:2020cjb} showed that corrections from residing in a local conformal cosmic void of $\sim20\%$ underdensity of a few tens of megaparsecs in diameter simultaneously solves the current tensions between measurements of the Hubble constant $H_0$ as well as of the amplitude of matter fluctuations $\sigma_8$ and restores a preference for a consistent CMB lensing amplitude of $A_L=1$ (Sec.~\ref{sec:intro}).
Specifically, in the frame adopted in Ref.~\cite{Bose:2020cjb}, the local CMB temperature $T_0$ deviates from that of the cosmological background $\bar{T}_0$ as $T_0 = \bar{T}_0(1+\delta_0)^{1/3}$, which changes parameters inferred from the CMB, particularly lowering $\sigma_8$ and $A_L$ and increasing $H_0$.
Allowing for spatial curvature, Planck, baryon acoustic oscillations and $H_0$ data then prefer an open and hotter universe at $>3\sigma$~\cite{Bose:2020cjb}.

The framework established in Secs.~\ref{sec:gravity}--\ref{sec:backgrounds} can be used to study further properties and observable signatures of conformal inhomogeneities.
In the following, Sec.~\ref{sec:confvoid} will focus on their nonlinear evolution whereas Sec.~\ref{sec:linear} will explore the linear limit, where observable signatures will be highlighted for both regimes.

\subsubsection{Conformal cosmic void} \label{sec:confvoid}

Consider a local conformal top-hat inhomogeneity with scale factor $\hat{a}$ of a smooth FLRW background metric with scale factor $a$.
Following Sec.~\ref{sec:nonlinearities}, this defines the conformal factor $\psi = \hat{a}/a=(1+\delta)^{-1/3}$.
Using the second Friedmann equation~\eqref{eq:ii} both for $\hat{a}$ and $a$, one recovers the spherical collapse equation,
\beq
    \ddot{\delta} + 2H\dot{\delta} - \frac{4}{3}\frac{\dot{\delta}^2}{1+\delta} = \frac{\kappa^2c^4}{2} \bar{\rho}_m (1+\delta)\delta \,, \label{eq:sphcoll}
\eeq
where dots denote derivatives with respect to the cosmic time $t$ of the smooth FLRW background, $dt=a\,d\tau$.
Note that while conformal times between the two frames agree, cosmic times do not, $t\neq\hat{t}$.
The evolution of $\delta$ and thus of $\hat{a}$ can be solved for setting the initial conditions at $a_i\ll1$ in the matter-dominated regime, where $\delta \propto a$.
The interior of the conformal top-hat inhomogeneity thus evolves as a usual top-hat density contrast.
This is not surprising in light of Birkhoff's theorem.
Density perturbations can be viewed as separate universes with different background density and curvature.
For a top-hat density perturbation this becomes manifest adopting Conformal Fermi Coordinates~\cite{Dai:2015jaa}.
Also recall from Sec.~\ref{sec:static} that any smooth FLRW metric describing the interior of the top hat has vanishing Weyl tensor and is thus conformally flat.

Besides the impact of a local conformal void on currently measured cosmological parameter tensions, the presence of conformal perturbations can have further observable implications.
Revisiting the relation of measured redshift $z$ to its background counterpart $\bar{z}$ in Sec.~\ref{sec:nonlinearities},
\beq
    1 + z = \left(\frac{1+\delta_s}{1+\delta_0}\right)^{1/3} (1 + \bar{z}) \,,
\eeq
one can see that redshifts are not only impacted by the environment of the observer but also by that of the sources, where both satisfy the evolution equation~\eqref{eq:sphcoll}.
Particularly, this implies that photons emitted from galaxy clusters ($\delta_s>0$) and detected in our underdense environment ($\delta_0<0$) may be significantly more redshifted than expected from $\bar{z}$.
As a consequence, one may attribute larger distances and older ages to observed galaxies in clusters, and consequently find larger masses and population sizes for those than expected.
This effect may offer an explanation for recent observations~\cite{Naidu:2022, Castellano:2022, Robertson:2022gdk}.
Of course, the local underdense environment may also lead to astronomical signatures in our immediate neighbourhood, and it is worthwhile noting that measurements of nearby galaxy groups~\cite{Karachentsev:2018ysz}, clusters~\cite{Bohringer:2019tyj} and peculiar velocities~\cite{Tully:2019ngb} indeed seem to indicate such an underdense local region.

In view of these effects, one may wonder about the nature of the coupling of matter fields with the local scale factor at a source $\psi = a_s$ suggested by Eqs.~\eqref{eq:psiaction}--\eqref{eq:energy}.
After all, photon-emitting sources typically lie in virialised regions, which have decoupled from the cosmic expansion, or inside compact bodies, and so $a_s = \mathrm{const}.$ at the immediate location of the emitting particle process.
Likewise, $a_0=\mathrm{const}.$ at the immediate position of our detectors.
However, when observing sources from virialised regions that are detached from ours, an expansion takes place between them.
In the non-expanding frame, the coupling hence corresponds to an effective coupling of the entire virialised region to the evolving scale factor of its environment and of that to the cosmological background or the space separating this region from ours.
This coupling can then be viewed as acting on the individual particles composing those regions.
Schematically, one has $\psi = a = a_{vir} \cdot a_{env} \cdot \bar{a}$, where $a_{vir} = \mathrm{const}.$ and $a_{env}$ may represent a composition of environments, $a_{env} = a_{env(1)} \cdot\cdot\cdot a_{env(n)}$.

\subsubsection{Linear cosmological perturbations} \label{sec:linear}

To explore further observable signatures attributed to the presence of conformal inhomogeneities in the Cosmos, one may resort to linear perturbation theory.
For that, consider the linearised conformal factor $\psi^2(\tau,\mathbf{x}) = 1 + 2\delta \psi(\tau,\mathbf{x})$, where $\left|\delta \psi\right|\ll1$, acting in front of the linearly perturbed FLRW line element in conformal Newtonian (or longitudinal) gauge, Eq.~\eqref{eq:linmetric}.
Operating in the linear regime, one may simply absorb the conformal factor by defining the new potentials $\hat{\Psi} \equiv \Psi+\delta \psi$ and $\hat{\Phi} \equiv \Phi+\delta \psi$.
The usual linearly perturbed Einstein field equations apply for $\hat{\Psi}$ and $\hat{\Phi}$.
Importantly, however, the notion of cosmic time may change with the absorption of the conformal factor, $dt = a\,d\tau \neq a(1+\delta\psi)d\tau$.
Given linearity, one may split the perturbed energy-momentum tensor as $\delta T_{\mu\nu} = \delta T_{\mu\nu}^{(\psi)} + \delta T_{\mu\nu}^{(0)}$, where $\delta T_{\mu\nu}^{(0)}$ determines $\Psi$ and $\Phi$.
Thus, the linearised field equations for $\delta\psi$ become
\beqa
     \nabla^2 \delta\psi -\frac{3a^2H}{c^2} \left ( \dot{\delta\psi} - H \delta\psi \right) & = & -\frac{\kappa^2c^2}{2}a^2 \delta\rho^{(\psi)} \,, \label{eq:lin1} \\
     \nabla^2\dot{\delta \psi} - H \nabla^2\delta\psi & = & \frac{\kappa^2c^2}{2} a\left(1+w\right)\bar{\rho}\theta^{(\psi)} \,, \label{eq:lin2} \\
     \nabla^2\delta\psi - \frac{3a^2}{2c^2} \left[\ddot{\delta\psi} + 2H\dot{\delta\psi} - \left( H^2 + 2\frac{\ddot{a}}{a}\right) \delta\psi \right] & = &  \frac{3\kappa^2c^2}{4} a^2 \delta P^{(\psi)} \,, \label{eq:lin3} \\ 
    \nabla^2 \delta\psi & = & \frac{3\kappa^2c^2}{4}a^2\left(1+w \right) \bar{\rho} \sigma^{(\psi)} \,, \label{eq:lin4}
\eeqa
where the density perturbation $\delta\rho^{(\psi)} = \bar{\rho}\delta^{(\psi)}$, the divergence of the fluid velocity $\theta^{(\psi)}$, the pressure perturbation $\delta P^{(\psi)}$, and the shear (or anisotropic) stress $\sigma^{(\psi)}$ describe the fluid giving rise to the conformal inhomogeneity.
Energy-momentum conservation implies
\beqa
    \dot{\delta}^{(\psi)} & = & -(1+w) \left( \frac{\theta^{(\psi)}}{a} + 3 \dot{\delta\psi} \right) - 3 H \left( \frac{c_s^{2(\psi)}}{c^2} - w \right) \delta^{(\psi)} \,, \\
    \dot{\theta}^{(\psi)} & = & -H(1-3w)\theta^{(\psi)} - \frac{\dot{w}}{1+w}\theta^{(\psi)} - \frac{1}{a} \left( \frac{c_s^{2(\psi)}}{1+w} \nabla^2 \delta^{(\psi)} - c^2\nabla^2\sigma^{(\psi)} + c^2\nabla^2\delta^{(\psi)} \psi \right) \,,
\eeqa
where $c_s^{2(\psi)} = c^2\delta P^{(\psi)} / \delta \rho^{(\psi)}$ denotes the sound speed squared.

It should be emphasised again that dots represent here derivatives with respect to the cosmic time $dt = a \, d\tau$ rather than $dt = a(1+\delta\psi)d\tau$, which can give rise to effective fluid perturbations.
To illustrate this point, consider separating off a fraction of the scale factor of the smooth FLRW metric as $a \rightarrow a(1+\delta a)$ such that $\delta\psi = \delta a(t)$ and $\nabla^2\delta a = 0$ everywhere.
Eqs.~\eqref{eq:lin1} and \eqref{eq:lin3} then directly follow from the linearised first and second Friedmann equations, Eqs.~\eqref{eq:00} and \eqref{eq:ii}, respectively, under the transformations of the scale factor and of cosmic time, where $\delta\rho^{(a)}$ and $\delta P^{(a)}$ are the density and pressure perturbations attributed to $\delta a$.
For instance, one may separate out a different background matter density $\bar{\rho}_m^{(a)}$ (see Sec.~\ref{sec:gravity}), giving rise to an effective density contrast $\delta\rho^{(a)}_m = \bar{\rho}^{(a)}_m \delta^{(a)}$ and $\delta P^{(a)}_m = 0$.
As another example, given that spatially curved FRLW metrics are conformally flat (Sec.~\ref{sec:static}), one may treat a small spatial curvature as a linear conformal perturbation $\delta \psi = \delta a$ of a flat background.

From the field equations~\eqref{eq:lin1}--\eqref{eq:lin4}, one can already identify some interesting observable implications of linear conformal perturbations without the need for solving the system of differential equations.
Consider the fluid comoving density in Fourier space,
\beq
    \Delta = \delta + 3(1+w) \frac{aH}{c^2k^2} \theta \,.
\eeq
Note that for $\delta \psi = \delta a (t)$, this density vanishes, $\Delta^{(a)} = 0$.
Eqs.~\eqref{eq:lin1}, \eqref{eq:lin2} and \eqref{eq:lin4} more generally imply
\beq
    k^2\delta\psi = \frac{\kappa^2c^2}{2}a^2\Delta^{(\psi)} = -\frac{3\kappa^2c^2}{4}a^2\left(1+w\right)\bar{\rho} \sigma^{(\psi)} \,.
\eeq
For a matter perturbation, one has $2k^2\Phi = \kappa^2c^2a^2\bar{\rho}_m \Delta_m$ such that the total potential satisfies
\beq
    k^2 \hat{\Phi} = \frac{\kappa^2c^2}{2} a^2 \bar{\rho}_m \left( \Delta_m + \Delta^{(\psi)} \right) \,.
\eeq
Since $\sigma_m=0$, the potentials relate as $\Psi=-\Phi$ and
\beq
    -k^2 \hat{\Psi} = \frac{\kappa^2c^2}{2} a^2 \bar{\rho}_m \left( \Delta_m - \Delta^{(\psi)} \right) \,. \label{eq:modpsi}
\eeq
For the lensing potential, one finds
\beq
    \frac{k^2}{2} \left( \hat{\Phi}-\hat{\Psi} \right) = \frac{\kappa^2c^2}{2} a^2 \bar{\rho}_m \Delta_m \,.
\eeq
Thus, lensing remains unaffected by the presence of the conformal inhomogeneity, as expected for a conformal factor, whereas dynamic, or kinematic, effects such as clustering are modified in Eq.~\eqref{eq:modpsi}.
The conformal inhomogeneity acts like an effective modification of gravity with
\beqa
    \mu(a,k) & \equiv & -\frac{2 k^2 \hat{\Psi}}{\kappa^2c^2 a^2 \bar{\rho}_m \Delta_m} = 1 - \frac{\Delta^{(\psi)}}{\Delta_m} \,, \label{eq:mu} \\
    \Sigma(a,k) & \equiv & -\frac{k^2\left( \hat{\Phi}-\hat{\Psi} \right)}{\kappa^2c^2 a^2 \bar{\rho}_m \Delta_m} = 1 \,.
\eeqa
Thus, for clusters with $\Delta_m > 0$ and for $\Delta^{(\psi)} < \Delta_m$, this implies that cluster masses inferred dynamically, or kinematically, $M_{dyn}$, are underestimated compared to masses inferred from lensing, $M_{len}$.
This mass bias is indeed observed but typically associated with deviations from hydrostatic equilibrium affecting the estimations of $M_{dyn}$~\cite{Planck:2013lkt}.
The growth suppression $\mu<1$ in Eq.~\eqref{eq:mu} may also act to lower $\sigma_8$ as preferred by galaxy surveys~\cite{Heymans:2020gsg}.
Note that in contrast, modified gravity models typically produce $\mu>1$ as well as $M_{dyn} > M_{len}$~\cite{Terukina:2013eqa} for theoretical stability reasons~\cite{Kennedy:2018gtx}.
It is also worthwhile noting that for conformal inhomogeneities, $\Delta^{(\psi)}/\Delta_m$ may also be negative, for example in voids, such that $\mu>1$ and $M_{dyn} > M_{len}$ instead.

\subsection{Dark matter, inflation and baryogenesis} \label{sec:dm}

In Sec.~\ref{sec:dS}, an interpretation of dark matter as the effect of an early-time convergent increase of particle masses was given.
Similarly, as discussed in Sec.~\ref{sec:EdS}, dark energy could be interpreted as the late-time divergent increase of particle masses.
These interpretations were motivated by the change of the cosmic geometry into de Sitter and Einstein-de Sitter space, respectively.
In the following, another candidate of geometric origin for dark matter as well as for inflation shall briefly be explored, which is motivated from treating the scale factor as a scalar field as in Sec.~\ref{sec:cosmologicalconstant}.

Let us revisit the action~\eqref{eq:scalfieldaction}, which for the field $\chi \equiv 1 + 2\delta\psi$ becomes
\beq
    S_{\chi} = \int d^4x \sqrt{-g} \psi_*^2\left[ - \frac{1}{2} \partial_{\mu}\chi \partial^{\mu}\chi - \frac{1}{2} m_{\chi}^2\chi^2 \right] \,, \label{eq:Schi}
\eeq
where $m^2_{\chi} = 4\Lambda\psi_*^2/3$ and $S_{\chi}\simeq4S_{\psi}$.
Instead of the Minkowski space in Sec.~\ref{sec:cosmologicalconstant}, however, $g_{\mu\nu}$ shall describe an evolving smooth FLRW background obtained after a conformal transformation, for example, of a conformal inhomogeneity similar to Sec.~\ref{sec:nonlinearities}.
Thus, $\delta\psi$ describes the perturbation of the relative scale factor around $\psi_* = a_*/\bar{a}_*$.
In the interior, the field is only time dependent such that the scalar field equation implies
\beq
    \ddot{\chi} + 3H\dot{\chi} + m_{\chi}^2\chi = 0 \,. \label{eq:axion}
\eeq
Eq.~\eqref{eq:axion} suggests that $\chi$ may act as an axion~\cite{Marsh:2015xka}.
Let us assume the initial condition $\dot{\chi}=0$.
For $H > m_{\chi}$, the axion field is overdamped and frozen to its initial value with an equation of state of $w_{\chi}=-1$ such that it may act as dark energy or drive inflation.
For $H < m_{\chi}$, the axion field is underdamped, oscillating around $w_{\chi}=0$ with energy density $\rho_{\chi} \sim a^{-3}$.
Hence, in this case, it may act as dark matter.

Finally, the transformed action~\eqref{eq:psiaction} could also have implications for baryogenesis, irrespective of the axion treatment, where couplings could introduce a CP violation causing baryogenesis~\cite{Marsh:2015xka}.
For this, consider quantum gravity corrections in the action such as the term $R^2$.
Besides being a potential source of inflation~\cite{Starobinsky:1979ty}, this term can also provide a coupling relevant to baryogenesis in the conformally transformed action.
Specifically, with Eq.~\eqref{eq:Rconf} one finds from integration by parts the term
\beq
    \sqrt{-g}R^2 \propto -\sqrt{-\tilde{g}} \tilde{J}^{\mu} \partial_{\mu}\tilde{R} + \ldots \,, \label{eq:baryogenesis}
\eeq
defining the current $\tilde{J}^{\mu} \equiv \partial^{\mu}\ln \psi = \tilde{T}^{-1} \tilde{\nabla}^{\sigma} \tilde{T}^{\mu}_{\ \sigma}$ with Eq.~\eqref{eq:energy}, where $\tilde{\nabla}_{\mu} \tilde{J}^{\mu}$ can be non-vanishing with Eq.~\eqref{eq:scal}.
 This is reminiscent of the CP-violating coupling $\sqrt{-g}J_{B-L}^{\mu}\partial_{\mu}R$ for gravitational baryogenesis~\cite{Davoudiasl:2004gf, Hook:2014mla, Boudon:2020qpo}, where for a non-conserved baryon-lepton current $J_{B-L}$ net $B-L$ charge is generated.

 A more detailed analysis of Eqs.~\eqref{eq:axion} and \eqref{eq:baryogenesis} is left for future work.

\section{Conclusions} \label{sec:conclusions}

Despite the resounding success of the standard cosmological model ($\Lambda$CDM) in reproducing our wealth of cosmological observations, there remains a number of unexplained larger and smaller tensions among the cosmological model parameters inferred from our different data sets.
Besides these observational tensions, significant gaps persist in our theoretical understanding of key $\Lambda$CDM ingredients.
The cosmological constant problem, the nature of dark matter and inflation, or the origin of the matter-antimatter asymmetry in the Universe constitute particularly important and profound enigmas among them.
Both these observational and theoretical challenges motivate the development and search for new physics.
In contrast, a less radical approach to venturing beyond the Standard Model is the simple mathematical reformulation of our theoretical frameworks underlying it.
This can offer reinterpretations and possibly even solutions of these problems.

This article explored the implications of casting cosmology into different spacetime geometries by a combination of conformal transformations and the dimensional reduction of the cosmological spacetime metric applied to the Einstein-Hilbert action and Einstein field equations.
Although simply a mathematical manipulation that leaves physical measurements unaffected, new perspectives on the observational and theoretical challenges to $\Lambda$CDM were attained.
A particular focus was thereby placed on casting cosmology into Minkowski space.
Clearly, the Minkowski metric occupies a special place among the metrics one could transform the cosmic geometry into.
It is static and flat as well as the spacetime of special relativity and quantum field theory or indeed of the entire Standard Model.

Applications of the formalism developed here are manifold, and the article has identified a number of those, which shall briefly be summarised in the following.
It should be emphasised, however, that the discussions provided in this context were far from exhaustive and are meant to serve as inspiration for further work on Minkowskian cosmology.

(1)~Rather than due to an expanding Cosmos, observed redshifts to distant galaxies can be interpreted as the evolution of particle masses after conformal transformation of the FLRW metric into static or Minkowski space.
This recovers the Friedmann equations and the evolution of energy densities, and importantly, leaves measurements such as redshifts invariant.

(2)~The transformation into Minkowski space can be performed even in the presence of spatial curvature.
This can either be achieved with a conformal transformation adopting suitable coordinates in which the vanishing Weyl tensor and thus the conformal flatness of the spatially curved FLRW metric becomes apparent or through the combination of dimensional reduction and conformal transformations of the metric.
In this case, spatial curvature can be interpreted as varying length scale.
Alternatively, small spatial curvature may also be treated as an example of a linear conformal inhomogeneity.

(3)~Following a similar procedure to the decomposition of the spatially curved FLRW metric, cosmic structure and gravitational waves can be viewed as distortions and oscillations of mass, length, and time scales in the Minkowski frame.
Alternatively, these distortions may be interpreted as the variation of the speed of light, the Planck constant and the gravitational coupling.

(4)~A particularly promising application of the framework was identified with casting the cosmological constant problem into Minkowskian cosmology.
Specifically, in the Minkowski frame, $\Lambda$ was found to take on a natural value set by the Planck scale and the size of the present Universe.
This value is furthermore consistent with the gravitating vacuum ground state energy density obtained from canonical quantisation of the scale factor with cutoff at the evolving Planck mass.
This reduces the cosmological constant to the \emph{Why Now?}~or coincidence problem, for which anthropic arguments can be invoked as remedy.
This can be in form of a simple observer's bias for our location in cosmic history that does not necessitate a multiverse.

(5)~Geometric candidates for dark matter and dark energy also naturally emerge from the formalism.
For example, adopting an Einstein-de Sitter and de Sitter background suggested an interpretation of dark energy and dark matter as evolving particle masses that respectively diverge and converge with proceeding age of the Universe.
Another geometric dark matter candidate arises from treating conformal inhomogeneities in the scale factor as a scalar field.
This field can behave as an axion and could act as dark matter or dark energy.

(6)~Geometric candidates for inflation and baryogenesis constitute another application.
Besides serving as dark sector candidate, the geometric axion field also offers a promising candidate for inflation.
Couplings to the field may furthermore introduce a CP violation that could give rise to baryogensis.
Another interesting candidate is found in the Minkowski frame of a possible $R^2$ term in the Einstein-Hilbert action that can arise from quantum gravity corrections.
Besides offering a source for inflation, the transformed action contains a term that resembles that of the CP-violating coupling of gravitational baryogenesis.

(7) Finally, the presence of conformal inhomogeneities in our Universe can offer a simultaneous solution to the Hubble tension, the discrepant measurements in the amplitude of matter fluctuations as well as the preference for a non-standard lensing amplitude in CMB anisotropies, provided we are located in an underdense conformal region of space.
Conformal inhomogeneities can furthermore enhance redshifts to distant galaxies, causing larger distances and older ages to be attributed to galaxies in clusters and consequently larger masses and population sizes than otherwise expected for those.
Another effect is a mass bias for the inferred masses of galaxy clusters, where masses inferred from lensing are larger than those inferred dynamically or kinematically.
It is worth highlighting that these predictions agree qualitatively with trends identified in current observations and may thus be worthwhile further investigating.

There is good reason to expect that this list of applications is far from exhaustive.
For instance, one may also employ the formalism for the conformally flat interior of black holes and stars to gain a new perspective on the physical nature of those.
Moreover, for a Schwarzschild or Schwarzschild-de Sitter black hole, one can perform the analogous change of coordinates to Eqs.~\eqref{eq:curvedFLRW}--\eqref{eq:curvedFLRWtransformed} such that their metrics assume the form of Eq.~\eqref{eq:metric}.
From Eq.~\eqref{eq:prescales} it then follows that $\tilde{m} = m \sqrt{1-r_s/r - \Lambda r^2/3}$ for the mass of a test particle, which vanishes at the horizon $r_s$ for $\Lambda=0$.
The changes of time and length scales can be worked out accordingly.
Further applications may include the exploration of the Minkowski picture in the Palatini formalism or its generalisation to teleparallel gravity~\cite{Bahamonde:2021gfp, Bernardo:2021bsg} and other extended theories of gravity~\cite{Bull:2015stt, Joyce:2016vqv}.
The discussion provided here on the applications of the formalism only served as brief adumbration of those, intended as a basis for further work on Minkowskian cosmology.
The promising implications highlighted in this context certainly suggest that this will be a worthwhile endeavour.

\section*{Acknowledgements}

The author thanks Charles Dalang and Gaëtan Brunetto for useful discussions.
This work was supported by a Swiss National Science Foundation (SNSF) Professorship grant (Nos.~170547 \& 202671).


\bibliographystyle{JHEP}
\bibliography{minkowski}

\providecommand{\href}[2]{#2}\begingroup\raggedright\begin{thebibliography}{10}

\bibitem{Weinberg:1988cp}
S.~Weinberg, {\it {The Cosmological Constant Problem}},  {\em Rev. Mod. Phys.}
  {\bf 61} (1989) 1--23.

\bibitem{Martin:2012bt}
J.~Martin, {\it {Everything You Always Wanted To Know About The Cosmological
  Constant Problem (But Were Afraid To Ask)}},  {\em Comptes Rendus Physique}
  {\bf 13} (2012) 566--665, [\href{http://arxiv.org/abs/1205.3365}{{\tt
  arXiv:1205.3365}}].

\bibitem{Bertone:2016nfn}
G.~Bertone and D.~Hooper, {\it {History of dark matter}},  {\em Rev. Mod.
  Phys.} {\bf 90} (2018), no.~4 045002,
  [\href{http://arxiv.org/abs/1605.04909}{{\tt arXiv:1605.04909}}].

\bibitem{Baumann:2009ds}
D.~Baumann, {\it {Inflation}},  in {\em {Theoretical Advanced Study Institute
  in Elementary Particle Physics}: {Physics of the Large and the Small}},
  pp.~523--686, 2011.
\newblock \href{http://arxiv.org/abs/0907.5424}{{\tt arXiv:0907.5424}}.

\bibitem{Linde:2014nna}
A.~Linde, {\it {Inflationary Cosmology after Planck 2013}},  in {\em {100e
  Ecole d'Ete de Physique: Post-Planck Cosmology}}, pp.~231--316, 2015.
\newblock \href{http://arxiv.org/abs/1402.0526}{{\tt arXiv:1402.0526}}.

\bibitem{Riotto:1998bt}
A.~Riotto, {\it {Theories of baryogenesis}},  in {\em {ICTP Summer School in
  High-Energy Physics and Cosmology}}, pp.~326--436, 7, 1998.
\newblock \href{http://arxiv.org/abs/hep-ph/9807454}{{\tt hep-ph/9807454}}.

\bibitem{Canetti:2012zc}
L.~Canetti, M.~Drewes, and M.~Shaposhnikov, {\it {Matter and Antimatter in the
  Universe}},  {\em New J. Phys.} {\bf 14} (2012) 095012,
  [\href{http://arxiv.org/abs/1204.4186}{{\tt arXiv:1204.4186}}].

\bibitem{Riess:2021jrx}
A.~G. Riess et~al., {\it {A Comprehensive Measurement of the Local Value of the
  Hubble Constant with 1 km s$^{-1}$ Mpc$^{-1}$ Uncertainty from the Hubble
  Space Telescope and the SH0ES Team}},  {\em Astrophys. J. Lett.} {\bf 934}
  (2022), no.~1 L7, [\href{http://arxiv.org/abs/2112.04510}{{\tt
  arXiv:2112.04510}}].

\bibitem{Heymans:2020gsg}
C.~Heymans et~al., {\it {KiDS-1000 Cosmology: Multi-probe weak gravitational
  lensing and spectroscopic galaxy clustering constraints}},  {\em Astron.
  Astrophys.} {\bf 646} (2021) A140,
  [\href{http://arxiv.org/abs/2007.15632}{{\tt arXiv:2007.15632}}].

\bibitem{Aghanim:2018eyx}
{\bf Planck} Collaboration, N.~Aghanim et~al., {\it {Planck 2018 results. VI.
  Cosmological parameters}},  {\em Astron. Astrophys.} {\bf 641} (2020) A6,
  [\href{http://arxiv.org/abs/1807.06209}{{\tt arXiv:1807.06209}}].

\bibitem{Bull:2015stt}
P.~Bull et~al., {\it {Beyond $\Lambda$CDM: Problems, solutions, and the road
  ahead}},  {\em Phys. Dark Univ.} {\bf 12} (2016) 56--99,
  [\href{http://arxiv.org/abs/1512.05356}{{\tt arXiv:1512.05356}}].

\bibitem{Perivolaropoulos:2021jda}
L.~Perivolaropoulos and F.~Skara, {\it {Challenges for \ensuremath{\Lambda}CDM:
  An update}},  {\em New Astron. Rev.} {\bf 95} (2022) 101659,
  [\href{http://arxiv.org/abs/2105.05208}{{\tt arXiv:2105.05208}}].

\bibitem{Kunz:2007rk}
M.~Kunz, {\it {The dark degeneracy: On the number and nature of dark
  components}},  {\em Phys. Rev. D} {\bf 80} (2009) 123001,
  [\href{http://arxiv.org/abs/astro-ph/0702615}{{\tt astro-ph/0702615}}].

\bibitem{Lombriser:2015sxa}
L.~Lombriser and A.~Taylor, {\it {Breaking a Dark Degeneracy with Gravitational
  Waves}},  {\em JCAP} {\bf 03} (2016) 031,
  [\href{http://arxiv.org/abs/1509.08458}{{\tt arXiv:1509.08458}}].

\bibitem{Joyce:2016vqv}
A.~Joyce, L.~Lombriser, and F.~Schmidt, {\it {Dark Energy Versus Modified
  Gravity}},  {\em Ann. Rev. Nucl. Part. Sci.} {\bf 66} (2016) 95--122,
  [\href{http://arxiv.org/abs/1601.06133}{{\tt arXiv:1601.06133}}].

\bibitem{Jana:2020vov}
S.~Jana, C.~Dalang, and L.~Lombriser, {\it {Horndeski theories and beyond from
  higher dimensions}},  {\em Class. Quant. Grav.} {\bf 38} (2021), no.~2
  025003, [\href{http://arxiv.org/abs/2007.06907}{{\tt arXiv:2007.06907}}].

\bibitem{Endean:1997}
G.~{Endean}, {\it {Cosmology in Conformally Flat Spacetime}},  {\em
  Astrophysical Journal} {\bf 479} (Apr., 1997) 40--45.

\bibitem{Iihoshi:2007uz}
M.~Iihoshi, S.~V. Ketov, and A.~Morishita, {\it {Conformally flat FRW
  metrics}},  {\em Prog. Theor. Phys.} {\bf 118} (2007) 475--489,
  [\href{http://arxiv.org/abs/hep-th/0702139}{{\tt hep-th/0702139}}].

\bibitem{Ibison:2007dv}
M.~Ibison, {\it {On the conformal forms of the Robertson-Walker metric}},  {\em
  J. Math. Phys.} {\bf 48} (2007) 122501,
  [\href{http://arxiv.org/abs/0704.2788}{{\tt arXiv:0704.2788}}].

\bibitem{Gron:2011yi}
O.~Gron and S.~Johannesen, {\it {FRW Universe Models in Conformally Flat
  Spacetime Coordinates. I: General Formalism}},  {\em Eur. Phys. J. Plus} {\bf
  126} (2011) 28, [\href{http://arxiv.org/abs/1103.4743}{{\tt
  arXiv:1103.4743}}].

\bibitem{Domenech:2016yxd}
G.~Dom\`enech and M.~Sasaki, {\it {Conformal frames in cosmology}},  {\em Int.
  J. Mod. Phys. D} {\bf 25} (2016), no.~13 1645006,
  [\href{http://arxiv.org/abs/1602.06332}{{\tt arXiv:1602.06332}}].

\bibitem{Dalang:2019fma}
C.~Dalang and L.~Lombriser, {\it {Limitations on Standard Sirens tests of
  gravity from screening}},  {\em JCAP} {\bf 10} (2019) 013,
  [\href{http://arxiv.org/abs/1906.12333}{{\tt arXiv:1906.12333}}].

\bibitem{Wetterich:2013aca}
C.~Wetterich, {\it {Universe without expansion}},  {\em Phys. Dark Univ.} {\bf
  2} (2013) 184--187, [\href{http://arxiv.org/abs/1303.6878}{{\tt
  arXiv:1303.6878}}].

\bibitem{Postma:2014vaa}
M.~Postma and M.~Volponi, {\it {Equivalence of the Einstein and Jordan
  frames}},  {\em Phys. Rev. D} {\bf 90} (2014), no.~10 103516,
  [\href{http://arxiv.org/abs/1407.6874}{{\tt arXiv:1407.6874}}].

\bibitem{Lombriser:2019ahl}
L.~Lombriser, {\it {Consistency of the local Hubble constant with the cosmic
  microwave background}},  {\em Phys. Lett. B} {\bf 803} (2020) 135303,
  [\href{http://arxiv.org/abs/1906.12347}{{\tt arXiv:1906.12347}}].

\bibitem{Bose:2020cjb}
B.~Bose and L.~Lombriser, {\it {Easing cosmic tensions with an open and hotter
  universe}},  {\em Phys. Rev. D} {\bf 103} (2021), no.~8 L081304,
  [\href{http://arxiv.org/abs/2006.16149}{{\tt arXiv:2006.16149}}].

\bibitem{Visser:2015iua}
M.~Visser, {\it {Conformally
  Friedmann\textendash{}Lema\^\i{}tre\textendash{}Robertson\textendash{}Walker
  cosmologies}},  {\em Class. Quant. Grav.} {\bf 32} (2015), no.~13 135007,
  [\href{http://arxiv.org/abs/1502.02758}{{\tt arXiv:1502.02758}}].

\bibitem{Blanco:2020ipk}
D.~S. Blanco and L.~Lombriser, {\it {Exploring the self-tuning of the
  cosmological constant from Planck mass variation}},  {\em Class. Quant.
  Grav.} {\bf 38} (2021), no.~23 235003,
  [\href{http://arxiv.org/abs/2012.01838}{{\tt arXiv:2012.01838}}].

\bibitem{Dai:2015jaa}
L.~Dai, E.~Pajer, and F.~Schmidt, {\it {On Separate Universes}},  {\em JCAP}
  {\bf 10} (2015) 059, [\href{http://arxiv.org/abs/1504.00351}{{\tt
  arXiv:1504.00351}}].

\bibitem{Naidu:2022}
R.~P. {Naidu} et~al., {\it {Two Remarkably Luminous Galaxy Candidates at z
  {\ensuremath{\approx}} 10-12 Revealed by JWST}},  {\em Astrophysical Journal
  Letters} {\bf 940} (Nov., 2022) L14,
  [\href{http://arxiv.org/abs/2207.09434}{{\tt arXiv:2207.09434}}].

\bibitem{Castellano:2022}
M.~{Castellano} et~al., {\it {Early Results from GLASS-JWST. III. Galaxy
  Candidates at z 9-15}},  {\em Astrophysical Journal Letters} {\bf 938} (Oct.,
  2022) L15, [\href{http://arxiv.org/abs/2207.09436}{{\tt arXiv:2207.09436}}].

\bibitem{Robertson:2022gdk}
B.~E. Robertson et~al., {\it {Discovery and properties of the earliest galaxies
  with confirmed distances}},  \href{http://arxiv.org/abs/2212.04480}{{\tt
  arXiv:2212.04480}}.

\bibitem{Karachentsev:2018ysz}
I.~D. Karachentsev and K.~N. Telikova, {\it {Stellar and dark matter density in
  the Local Universe}},  {\em Astron. Nachr.} {\bf 339} (2018), no.~7-8
  615--622, [\href{http://arxiv.org/abs/1810.06326}{{\tt arXiv:1810.06326}}].

\bibitem{Bohringer:2019tyj}
H.~B\"ohringer, G.~Chon, and C.~A. Collins, {\it {Observational evidence for a
  local underdensity in the Universe and its effect on the measurement of the
  Hubble Constant}},  {\em Astron. Astrophys.} {\bf 633} (2020) A19,
  [\href{http://arxiv.org/abs/1907.12402}{{\tt arXiv:1907.12402}}].

\bibitem{Tully:2019ngb}
R.~B. Tully, D.~Pomarede, R.~Graziani, H.~M. Courtois, Y.~Hoffman, and E.~J.
  Shaya, {\it {Cosmicflows-3: Cosmography of the Local Void}},  {\em Astrophys.
  J.} {\bf 880} (2019), no.~1 24, [\href{http://arxiv.org/abs/1905.08329}{{\tt
  arXiv:1905.08329}}].

\bibitem{Planck:2013lkt}
{\bf Planck} Collaboration, P.~A.~R. Ade et~al., {\it {Planck 2013 results. XX.
  Cosmology from Sunyaev\textendash{}Zeldovich cluster counts}},  {\em Astron.
  Astrophys.} {\bf 571} (2014) A20, [\href{http://arxiv.org/abs/1303.5080}{{\tt
  arXiv:1303.5080}}].

\bibitem{Terukina:2013eqa}
A.~Terukina, L.~Lombriser, K.~Yamamoto, D.~Bacon, K.~Koyama, and R.~C. Nichol,
  {\it {Testing chameleon gravity with the Coma cluster}},  {\em JCAP} {\bf 04}
  (2014) 013, [\href{http://arxiv.org/abs/1312.5083}{{\tt arXiv:1312.5083}}].

\bibitem{Kennedy:2018gtx}
J.~Kennedy, L.~Lombriser, and A.~Taylor, {\it {Reconstructing Horndeski
  theories from phenomenological modified gravity and dark energy models on
  cosmological scales}},  {\em Phys. Rev. D} {\bf 98} (2018), no.~4 044051,
  [\href{http://arxiv.org/abs/1804.04582}{{\tt arXiv:1804.04582}}].

\bibitem{Marsh:2015xka}
D.~J.~E. Marsh, {\it {Axion Cosmology}},  {\em Phys. Rept.} {\bf 643} (2016)
  1--79, [\href{http://arxiv.org/abs/1510.07633}{{\tt arXiv:1510.07633}}].

\bibitem{Starobinsky:1979ty}
A.~A. Starobinsky, {\it {Spectrum of relict gravitational radiation and the
  early state of the universe}},  {\em JETP Lett.} {\bf 30} (1979) 682--685.

\bibitem{Davoudiasl:2004gf}
H.~Davoudiasl, R.~Kitano, G.~D. Kribs, H.~Murayama, and P.~J. Steinhardt, {\it
  {Gravitational baryogenesis}},  {\em Phys. Rev. Lett.} {\bf 93} (2004)
  201301, [\href{http://arxiv.org/abs/hep-ph/0403019}{{\tt hep-ph/0403019}}].

\bibitem{Hook:2014mla}
A.~Hook, {\it {Baryogenesis from Hawking Radiation}},  {\em Phys. Rev. D} {\bf
  90} (2014), no.~8 083535, [\href{http://arxiv.org/abs/1404.0113}{{\tt
  arXiv:1404.0113}}].

\bibitem{Boudon:2020qpo}
A.~Boudon, B.~Bose, H.~Huang, and L.~Lombriser, {\it {Baryogenesis through
  asymmetric Hawking radiation from primordial black holes as dark matter}},
  {\em Phys. Rev. D} {\bf 103} (2021), no.~8 083504,
  [\href{http://arxiv.org/abs/2010.14426}{{\tt arXiv:2010.14426}}].

\bibitem{Bahamonde:2021gfp}
S.~Bahamonde, K.~F. Dialektopoulos, C.~Escamilla-Rivera, G.~Farrugia, V.~Gakis,
  M.~Hendry, M.~Hohmann, J.~Levi~Said, J.~Mifsud, and E.~Di~Valentino, {\it
  {Teleparallel gravity: from theory to cosmology}},  {\em Rept. Prog. Phys.}
  {\bf 86} (2023), no.~2 026901, [\href{http://arxiv.org/abs/2106.13793}{{\tt
  arXiv:2106.13793}}].

\bibitem{Bernardo:2021bsg}
R.~C. Bernardo, J.~L. Said, M.~Caruana, and S.~Appleby, {\it {Well-tempered
  Minkowski solutions in teleparallel Horndeski theory}},  {\em Class. Quant.
  Grav.} {\bf 39} (2022), no.~1 015013,
  [\href{http://arxiv.org/abs/2108.02500}{{\tt arXiv:2108.02500}}].

\end{thebibliography}\endgroup

\end{document}